\documentclass[fleqn,usenatbib]{mnras}

\usepackage{newtxtext,newtxmath}

\usepackage[T1]{fontenc}

\DeclareRobustCommand{\VAN}[3]{#2}
\let\VANthebibliography\thebibliography
\def\thebibliography{\DeclareRobustCommand{\VAN}[3]{##3}\VANthebibliography}

\usepackage{graphicx}	
\usepackage{amsmath}	

\def\aaps{A\&AS}
\def\apj{ApJ}
\def\apjl{ApJL}
\def\mnras{MNRAS}
\def\pasa{PASA}
\def\pasp{PASP}
\def\aj{AJ}
\def\araa{ARA\&A}
\def\aap{A\&A}
\def\apjs{ApJS}

\def\deg{$^{\circ}$}

\newcommand{\ie}{{\it i.e.}}

\newcommand{\kms}{km~s$^{-1}$}
\newcommand{\kmsm}{km~s$^{-1}$~Mpc$^{-1}$}
\newcommand{\Ha}{$\rm H\alpha$}

\newcommand{\hi}{{H{\sc i}}}
\newcommand{\htwo}{H$_2$}

\newcommand{\rband}{{\em r}-band}

\newcommand{\whi}{$W_{50}$}

\newcommand{\about}{$\sim$}
\newcommand{\Msun}{M$_\odot$}

\newcommand{\Mhi}{$M_{\rm HI}$}

\newcommand{\Mst}{$M_\star$}

\newcommand{\rmax}{$R_{\rm max}$}

\newcommand{\rhi}{$R_{\rm HI}$}

\newcommand{\vhi}{$V_{\rm HI}$}
\newcommand{\vopt}{$V_{\rm OPT}$}
\newcommand{\vs}{$V_{\rm rot}/\sigma$}

\newcommand{\re}{$r_{\rm e}$}

\newcommand{\bct}{\vspace{-8 pt} \noindent}  
\newcommand{\bc}{\textcolor{magenta}}    

\title[SAMI-\hi]{SAMI-\hi: The \hi\ view of the \Ha\ Tully-Fisher relation and data release}

\author[B. Catinella et al.]
{Barbara Catinella,$^{1,2}$\thanks{barbara.catinella@uwa.edu.au} 
Luca Cortese,$^{1,2}$
Alfred L. Tiley,$^{1,2}$
Steven Janowiecki,$^{3}$
Adam B. Watts,$^{1,2}$
\newauthor
Julia J. Bryant,$^{2,4,5}$
Scott M. Croom,$^{2,4}$
Francesco d'Eugenio,$^{6,7}$
Jesse van de Sande,$^{2,4}$
\newauthor
Joss Bland-Hawthorn,$^{2,4,5}$
Amelia Fraser-McKelvie,$^{1,2}$
Samuel N. Richards,$^{4}$
Sarah M. Sweet,$^{2,8}$
\newauthor
Daniel J. Pisano,$^{9,10,11}$
Nickolas Pingel,$^{12,13}$
Rebecca A. Koopmann,$^{14}$
Dillion Cottrill,$^{15}$
and Meghan Hill$^{16}$ \\
$^{1}$International Centre for Radio Astronomy Research, The University of Western Australia, Crawley, WA 6009, Australia\\
$^{2}$ARC Centre of Excellence for All Sky Astrophysics in 3 Dimensions (ASTRO 3D), Australia\\
$^{3}$University of Texas, Hobby-Eberly Telescope, McDonald Observatory, TX 79734, USA\\
$^{4}$Sydney Institute for Astronomy (SIfA), School of Physics, The University of Sydney, NSW 2006, Australia\\
$^{5}$Australian Astronomical Optics, Astralis-USydney, School of Physics, University of Sydney, NSW 2006, Australia\\
$^{6}$Kavli Institute for Cosmology, University of Cambridge, Madingley Road, Cambridge, CB3 0HA, UK\\
$^{7}$Cavendish Laboratory - Astrophysics Group, University of Cambridge, 19 JJ Thomson Avenue, Cambridge, CB3 0HE, UK\\
$^{8}$School of Mathematics and Physics, University of Queensland, Brisbane, QLD 4072, Australia\\
$^{9}$Dept. of Physics \& Astronomy, West Virginia University, P.O. Box 6315, Morgantown, WV 26506, USA\\
$^{10}$Center for Gravitational Waves and Cosmology, West Virginia University, Chestnut Ridge Research Building, Morgantown, WV 26505, USA\\
$^{11}$Astronomy Department, University of Cape Town, Rondebosch 7700, South Africa\\
$^{12}$Research School of Astronomy and Astrophysics, The Australian National University, Canberra, ACT 2611, Australia\\
$^{13}$Department of Astronomy, The University of Wisconsin-Madison, 475 N. Charter Street, Madison, WI 53706, USA\\
$^{14}$Department of Physics \& Astronomy, Union College, Schenectady, NY, 12308, USA\\
$^{15}$Department of Physics and Astronomy, State University of New York at Stony Brook, Stony Brook, NY, USA\\
$^{16}$Department of Mechanical and Aerospace Engineering, Benjamin M. Statler College of Engineering and Mineral Resources, West Virginia University, \\
1306 Evansdale Drive, Morgantown, WV 26506-6106, USA\\
}

\date{Accepted XXX. Received YYY; in original form ZZZ}

\pubyear{2022}

\begin{document}
\label{firstpage}
\pagerange{\pageref{firstpage}--\pageref{lastpage}}
\maketitle

\begin{abstract}
We present SAMI-\hi, a survey of the atomic hydrogen content of 296 galaxies with integral field spectroscopy available from the SAMI Galaxy Survey. The sample spans nearly 4 dex in stellar mass ($M_\star = 10^{7.4}-10^{11.1}~ \rm M_\odot$), redshift $z<0.06$, and includes new Arecibo observations of 153 galaxies, for which we release catalogues and \hi\ spectra. We use these data to compare the rotational velocities obtained from optical and radio observations and to show how systematic differences affect the slope and scatter
of the stellar-mass and baryonic Tully-Fisher relations. Specifically, we show that \Ha\ rotational velocities measured in the inner parts of galaxies (1.3 effective radii in this work) systematically underestimate \hi\ global measurements, with \hi/\Ha\ velocity ratios that increase
at low stellar masses, where rotation curves are typically still rising and \Ha\ measurements do not reach their plateau. As a result, the \Ha\ stellar mass Tully-Fisher relation is steeper (when \Mst\ is the independent variable) and has larger scatter than its \hi\ counterpart. Interestingly, we confirm the presence of a small fraction of low-mass outliers of the \Ha\ relation that are not present when \hi\ velocity widths are used and are not explained by ``aperture effects''. These appear to be highly disturbed systems for which \Ha\ widths do not provide a reliable estimate of the rotational velocity. 
Our analysis reaffirms the importance of taking into account differences in velocity definitions as well as tracers used when interpreting offsets from the Tully-Fisher relation, at both low and high redshifts and when comparing with simulations.

\end{abstract}




\section{Introduction}\label{s_intro}

\begin{figure*}
\begin{center}
\includegraphics[width=17.5cm]{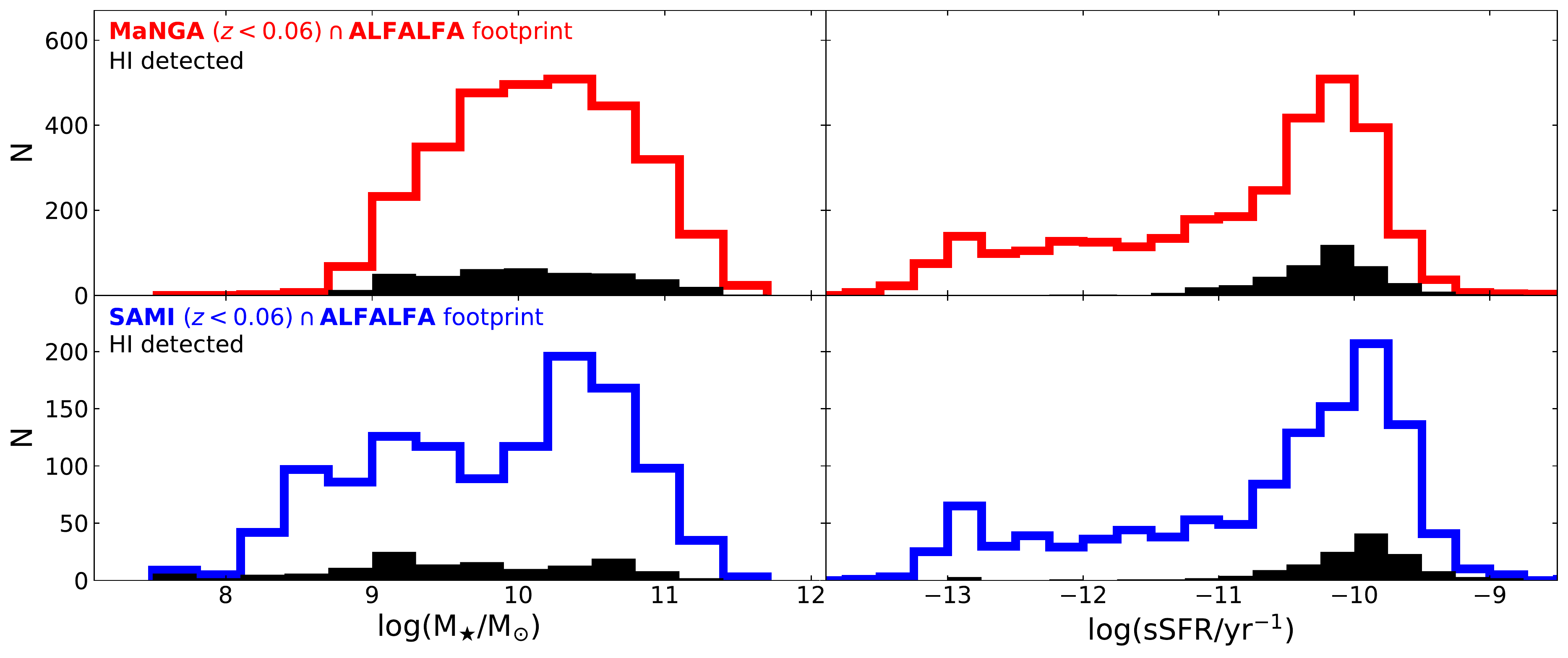}
\caption{{\it Top}: Stellar mass and specific SFR distributions of galaxies that lie in the overlap sky region between MaNGA and ALFALFA surveys, restricted to the ALFALFA redshift interval ($z<0.06$). Filled histograms show the subsets with \hi\ detections from ALFALFA.
{\it Bottom}: Same distributions for the SAMI Galaxy Survey, similarly restricted to ALFALFA footprint and redshift limits.
}
\label{fig_intro}
\end{center}
\end{figure*}

The past couple of decades have seen tremendous progress in our understanding of the physics of galaxies in the local Universe, thanks to the advent of 3D multiplexed large spectroscopic surveys in the optical and surveys of the cold interstellar medium at cm and mm wavelengths. The largest integral-field spectroscopy (IFS) surveys to date are the Sydney-AAO Multi-object Integral-field unit (SAMI, \citealp{sami,sami_dr3}) Galaxy Survey and the Mapping Nearby Galaxies at Apache Point Observatory (MaNGA, \citealp{bundy15}) survey, which combined provided detailed stellar and emission line diagnostics for over 13,000 galaxies at redshift $z<0.15$. Radio surveys of the cold gas component have measured the global content of atomic (\hi) and molecular hydrogen (\htwo, traced by carbon monoxide emission) for \about 50,000 and \about 1,000 galaxies at $z<0.06$ respectively, and mapped their kinematics and distributions down to the scales of giant molecular clouds, but only for small samples of very nearby galaxies (see \citealt{SC22_araa} and references therein).

The limited overlap currently present between cold gas and IFS samples prevents us from carrying out large statistical studies, which would allow us to characterize the physics of the gas-star formation cycle and quantify second-order trends that might drive the scatter of the main scaling relations.
Even when we focus on untargeted \hi\ surveys, which provided the vast majority of global \hi\ measurements currently available, the overlap between \hi\ {\it detections} and IFS surveys is very limited (largely due to differences in sky footprints and median redshift distributions) and significantly biased towards the most gas-rich, star-forming systems. This is illustrated in Fig.~\ref{fig_intro}, which shows the distributions of stellar mass and specific star-formation rate (SFR; specific $\rm SFR\equiv SFR/$\Mst) of MaNGA and SAMI, restricted to the sky footprint and redshift limit of the Arecibo Legacy Fast ALFA \citep[ALFALFA;][]{alfalfa,alfalfa100} untargeted \hi\ survey. 
Even within the ALFALFA volume,
only a small fraction of MaNGA and SAMI galaxies is detected in \hi\ by ALFALFA (\about 13\% in both cases, filled histograms), and these are all star-forming systems
(specific SFR $\gtrsim 10^{-11}$ $yr^{-1}$). 
This situation has motivated dedicated follow-up \hi\ observations of IFS samples, such as \hi-MaNGA \citep{himanga,stark21} and SAMI-\hi, which is presented in this paper.

Increasing the number of galaxies with global \hi\ and IFS observations is a step in the right direction, but there remain important differences between these datasets in terms of 
spatial and spectral resolution, as well as field-of-views. Thus, these observations are highly complementary and can provide important insights into how galaxies form and evolve, by shedding light on the interplay between their cold gas reservoirs and their stellar and ionised gas components at different spatial scales. For instance, \hi\ rotation curves and \Ha\ ones (obtained via Fabry-Perot, long-slit or IFS) have been combined to more accurately probe the galaxy gravitational potential for small samples \citep[e.g.,][]{noordermeer07,martinsson13,korsaga19}, but even the single velocity point provided by global \hi\ measurements can be useful to constrain the outer parts of rotation curves obtained from optical data \citep[e.g.,][]{cecil16,taranu17}. This synergy will become even more important when the next-generation \hi\ extragalactic surveys with the Square Kilometre Array \citep[SKA;][]{ska} and its  
pathfinder facilities (i.e., the Australian SKA Pathfinder, ASKAP: \citealt{askap}, \citealt{askap21}; the Karoo Array Telescope, MeerKAT: \citealt{meerkat}, \citealt{meerkat16}, and the APERture Tile in Focus upgrade of the Westerbork Synthesis Radio Telescope, APERTIF: \citealt{apertif}) will deliver data for millions of galaxies, most of which will be spatially unresolved or marginally resolved. For instance, the Widefield ASKAP L-band Legacy All-sky Blind surveY (WALLABY; \citealt{wallaby}) alone is expected to detect \about 210,000 galaxies out to $z\sim 0.1$ \citep{westmeier22}, whereas deeper \hi\ surveys such as DINGO
(Deep Investigation of Neutral Gas Origins, \citealt{dingo}), MIGHTEE-HI (MeerKAT International GHz Tiered Extragalactic Exploration, \citealt{mightee,mightee_hi}) and especially LADUMA (Looking At the Distant
Universe with the MeerKAT Array, \citealt{laduma12,laduma}), will detect (spatially unresolved) 21 cm emission from thousands of galaxies at higher redshifts.

With this in mind we carried out SAMI-\hi, a new \hi\ survey designed to increase the overlap between IFS and global \hi\ samples by targeting galaxies observed as part of the SAMI Galaxy Survey with the Arecibo radio telescope. Our goal is to provide the community with a sample that is representative of the star-forming main sequence population probed by SAMI to investigate how far we can push the combination of spatially unresolved \hi\ measurements and data from large IFS surveys. 
In this paper, we present the survey and \hi\ data release and compare dynamical scaling relations of local galaxies based on \hi\ velocity widths and \Ha\ rotation curves.  
In a companion paper \citep{watts22}, 
we use SAMI-\hi\ to investigate how IFS data give us insights into the physical origin of kinematic asymmetries observed in \hi\ global spectra, and whether \Ha\ and global \hi\ asymmetries are related.

One interesting area to explore with a sample with both IFS and \hi\ data available is that of kinematic scaling relations such as the Tully Fisher relation \citep[TFR;][]{TFR}.
The TFR is a tight, empirical correlation between the luminosity or stellar mass and the rotational velocity of disk galaxies, 
reflecting the underlying link between the luminous and total (dynamical) mass of galaxies.
The stellar TFR is known to break down at low stellar masses (\Mst $\lesssim 10^9$~\Msun), as gas-rich dwarf galaxies scatter below the relation (i.e., they are under-massive for their rotational velocity; \citealt{mcgaugh00}). A more fundamental relation is the {\it baryonic} TFR, which is obtained by replacing the stellar mass with the total baryonic mass
(i.e., the sum of stars and cold gas, \hi\ and molecular gas, \htwo) and remains linear in the dwarf regime (e.g., \citealt{mcgaugh00,begum08}; but see \citealt{mancera_pina19,mancera_pina20}, who found a small number of \hi-rich, ultradiffuse galaxies that scatter at low circular velocities for their baryonic mass). As a result, this relation has received significant attention in the past two decades \citep[e.g.,][]{gurovich04,mcgaugh05,geha06,stark09,zaritsky14,bradford16,lelli16,papastergis16,ponomareva18,lelli19,ponomareva21}.
Naturally, the samples available for baryonic TFR studies are smaller, since these require both optical and \hi\ measurements, with the 
\htwo\ content being usually neglected (because the interstellar medium of nearby galaxies is typically dominated by the \hi\ component; e.g., \citealt{SC22_araa}).
Rotational velocities are measured from \Ha\ or \hi\ emission using global spectra or rotation curves, or from stellar kinematics, and are often considered as proxies for circular velocities.
While this is a good approximation for cold gas in most cases (as the contribution from asymmetric drift for \hi\ and CO is generally small compared to their rotational velocities, except at very small stellar masses), deriving circular velocities from stellar or \Ha\ kinematics requires proper  correction for asymmetric drift using velocity dispersion measurements \citep[e.g.,][]{cresci09,leung18,varidel20}.

The TFR has been studied for decades by the \hi\ and optical communities,
initially for its value as a distance indicator and later for its importance for galaxy evolution studies \citep[e.g.,][]{aaronson79,bottinelli83,giovanelli97b,verheijen01,kannappan02,courteau07,pizagno07,reyes11}. More recently, IFS surveys (as well as molecular gas studies, e.g. \citealp{davis11,tiley16,topal18}) have contributed to the existing body of work (e.g., \citealp{luca14,bloom17,aquino18,barat19}). Interestingly, optical studies based on both IFS and long-slit spectroscopy have found an increase of scatter of the stellar-mass TFR at low stellar masses \citep{simons15,bloom17}, which is not seen in the \hi\ relation
(this scatter is in the opposite direction of the gas-rich dwarf galaxies that are outliers of the \hi\ stellar TFR, but not of the baryonic one). 
The scatter at low stellar mass of the optical TFR has been attributed to an increase of gas velocity dispersion \citep{simons15}, potentially associated with high \Ha\ kinematic asymmetry \citep{bloom17} in those systems.

Naturally, comparing \hi\ and optical and TFRs presents several challenges because, in addition to different sample selection, tensions could arise from the different kinematic tracers used (cold vs. ionised gas or stars), their radial extent, velocity definition (circular vs rotational) and the radius at which the velocity is measured \citep[see, e.g.,][]{courteau97,verheijen01,widths,lelli19}. \citet{lelli19} discussed some of these points in detail, and showed how these differences affect the baryonic TFR based on the Spitzer
Photometry and Accurate Rotation Curve \citep[SPARC;][]{SPARC} sample of galaxies with \hi\ and \Ha\ rotation curves available. 
This is particularly important when comparing 
TFR samples based on local galaxies with higher redshift measurements \citep[e.g.,][]{fernandez09,ubler17,tiley19} or predictions from theory and numerical simulations \citep[e.g.,][]{mmw98,navarro00,governato07,marinacci14,brook16,glowacki20}.
In this context, here we compare the stellar and baryonic mass TFRs based on these tracers and discuss the origin of the tensions mentioned above for a sample with both \Ha\ and \hi\ data available and uniformly measured.

This paper is organised as follows. In \S~\ref{s_sel} and \S~\ref{s_obs} we describe our sample selection and Arecibo observations. The kinematic measurements based on SAMI are summarised in \S~\ref{s_rcs}. In \S~\ref{s_res}, we characterise the SAMI-\hi\ sample and present our main results. We summarise and conclude in \S~\ref{s_concl}, and present the \hi\ data release in Appendix. All the distance-dependent quantities in this work are computed
assuming a cosmology with $H_0 = 70$ \kmsm, $\Omega_m= 0.3$ and $\Omega_\Lambda= 0.7$. We assume a \citet{chabrier03} initial mass function. AB magnitudes are used throughout the paper.

\section{Sample selection}\label{s_sel}

Our sample is selected from the SAMI Galaxy Survey \citep{sami,sami_dr3}, which includes 3068 galaxies with stellar masses greater than $\rm 10^7~M_\odot$ and redshifts $0.004 < z < 0.095$, observed with the Sydney-AAO Multi-object Integral field spectrograph \citep[SAMI;][]{sami_instr} on the 3.9-meter Anglo-Australian Telescope. The survey includes four volume-limited, stellar-mass selected galaxy samples extracted from the three equatorial fields of the Galaxy And Mass Assembly \citep[GAMA;][]{gama} survey. Additional fields covering 8 clusters probe higher density environments.
Each integral field unit (`hexabundle') covers a circular field of view of \about 14.9 arcsec in diameter. 13 hexabundles feed the SAMI spectrograph, which
provides a spectral coverage of the 3650--5800 and 6240--7450 \AA\ wavelength intervals with dispersions of 1.05 and 0.59 \AA\ pixel$^{-1}$, respectively. Individual SAMI fibers have a diameter of \about 1.6 arcsec each.

SAMI-\hi\ is the combination of two dedicated Arecibo proposals (see \S~\ref{s_obs}) and \hi\ detections from the ALFALFA survey. For our pilot proposal, we selected SAMI targets that had IFS data already available at the time  of the observations (2015) and were easily observable from Arecibo (i.e. in the GAMA equatorial fields, with declinations $\delta\geq 1$\deg\ and redshift $0.01<z<0.04$). For our second proposal, 
the sample selection was more clearly defined. Specifically, we started from the 1311 galaxies that had already been observed by SAMI in the GAMA regions in 2017. First, we imposed a declination ($0< \delta < 37$\deg) and redshift ($z<0.04$) cut to make Arecibo observations more efficient and avoid radio frequency interference (RFI). Second, we added a low stellar mass cut of 10$^{8.6}$~\Msun, to avoid objects whose \Ha\ velocity dispersion may be below the SAMI resolution limit and focused on star-forming galaxies, defined as objects with specific SFR
higher than 10$^{-11}$ $yr^{-1}$. 
Stellar masses are derived from the rest-frame $i$-band magnitude and $g-i$ colour as described in \cite{sami}. Integrated SFRs are taken from \cite{davies16} and were obtained via spectral energy distribution fitting using the MAGPHYS code \citep{magphys}. Lastly, we removed galaxies with effective radius ($r_{e}$) smaller than 2.5 arcsec (i.e., for which the inner regions would not be resolved) or larger than 7.5 arcsec (i.e., for which the SAMI coverage does not reach more than 1 effective radius). Effective radii were estimated from single Sersic fit to the Sloan Digital Sky Survey \citep[SDSS;][]{sdss} $r$-band images as described in \cite{kelvin12}. Excluding galaxies already detected by ALFALFA, this gave us a sample of 104 targets.

We obtained \hi\ emission-line spectra for 153 galaxies with these dedicated follow-up observations (78 from the first proposal and 75 from the second one). Together with 143 good-quality ALFALFA \hi\ detections (i.e., classified as ``code 1'', with a typical signal-to-noise ratio SNR$\geq 6.5$), our sample includes 296 galaxies with IFS from SAMI and global \hi\ spectra from Arecibo.

\begin{figure*}
\begin{center}
\includegraphics[width=17cm]{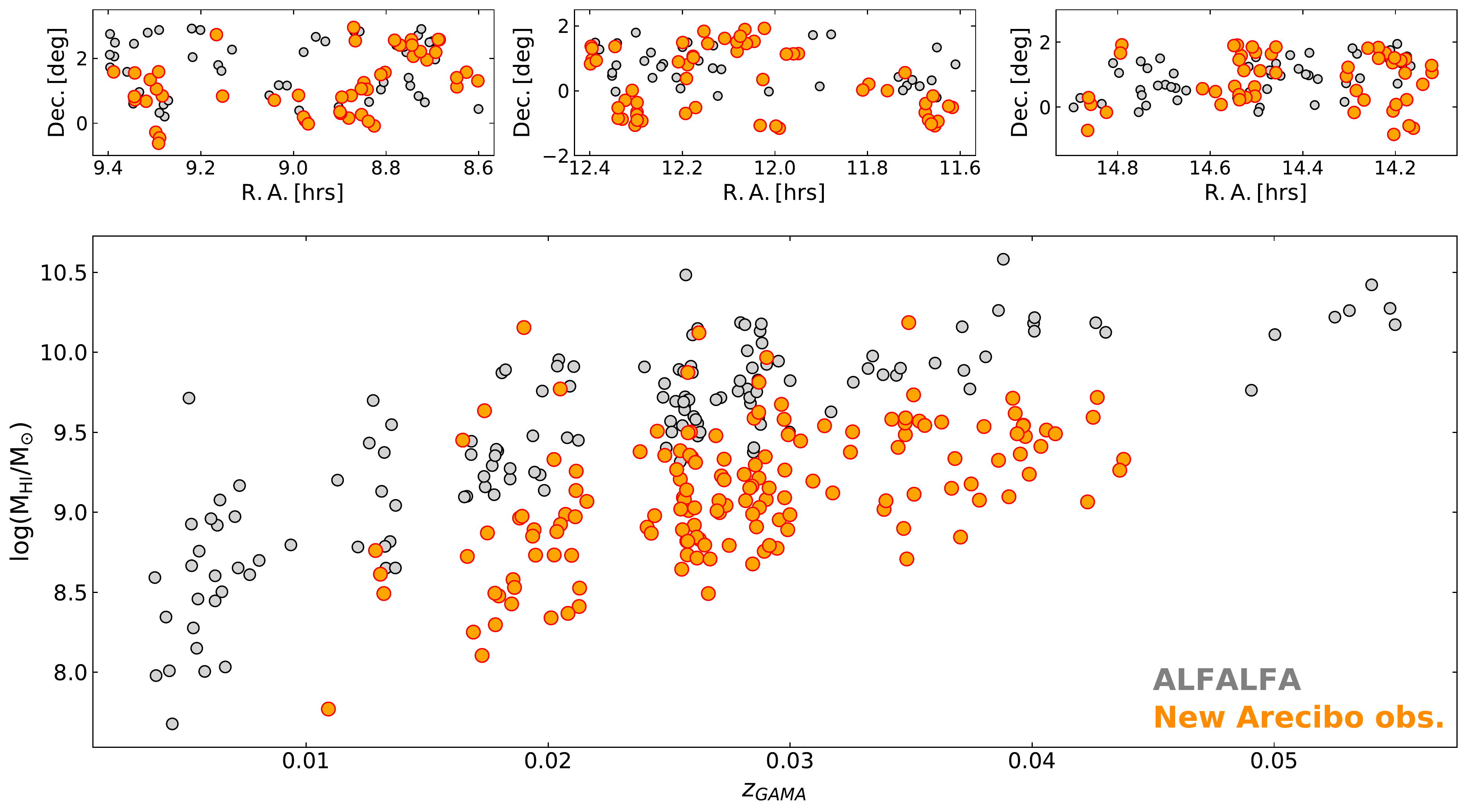}
\caption{Sky distribution (top) and plot of the \hi\ mass as a function of optical spectroscopic redshift (bottom) for the galaxies in SAMI-\hi. Orange and gray circles indicate the new Arecibo observations and ALFALFA data, respectively (see text).
}
\label{fig1}
\end{center}
\end{figure*}

\section{Arecibo observations and data reduction}\label{s_obs}

SAMI-\hi\ observations were carried out with the Arecibo radio telescope between 2015 and 2017. These were scheduled in 125 observing runs under programs A2934 and A3116; the total telescope time allocation was 161 hours. 

All the observations were carried out remotely except for one week in March 2017, when members of the  Undergraduate ALFALFA Team observed on site. We used the standard position-switching mode, the L-band wide receiver and the interim correlator as a backend. Two correlator boards with 12.5 MHz bandwidth, one polarization, and 2048 channels per spectrum (yielding a velocity resolution of 1.4 \kms\ at 1370 MHz before smoothing) were centered at or near the frequency corresponding to the GAMA redshift
of the target. The \hi\ spectra were recorded every second with 9-level sampling.
We observed each galaxy until detected, but moved to another target if there was no
hint of \hi\ signal within the first 4-5 on-off pairs. On-source integration times
ranged between 2 and 50 minutes per galaxy, with an average of 15 minutes. This is significantly deeper than ALFALFA, for which the integration time is \about 1 minute.

The data reduction, performed in the IDL environment, was the same as for the extended GALEX Arecibo SDSS Survey \citep[xGASS;][]{xgass}, and included Hanning smoothing, spectral bandpass subtraction, RFI excision, and flux calibration. Our RFI excision technique is illustrated in detail in \citet{highz}.
Briefly, the spectra obtained from each on/off pair are weighted by 1/$rms^2$,
where $rms$ is the root mean square noise measured in the signal-free portion of the spectrum, and co-added. The two orthogonal linear polarizations (kept separated up to this point) are averaged to produce the final spectrum; polarization mismatch, if significant, is noted in Appendix~B. The spectrum is then boxcar smoothed, spectral
baseline subtracted and measured as explained in \citet{gass1}. 

Observed velocity widths, \whi, are measured at the 50\% level of each flux peak.
We determine the peak flux density on both sides of the \hi\ profile and fit a straight line to each side (between the 15\% and 85\% levels of $f_p-rms$, where $f_p$ is the corresponding peak flux density). We 
identify the velocities $cz_r$, $cz_a$ corresponding to the
50\% peak flux density (from the fits) on the receding and approaching sides,
respectively, and define \whi\ as the difference between these two velocities (and the \hi\ redshift as their average). This is the same algorithm used to measure \hi\ velocity widths for large \hi\ data sets such as \citealt{springob05} (see their Section 3.2 for an illustration of this technique), ALFALFA \citep{alfalfa100} and xGASS.
The error on the observed velocity width is the sum in quadrature of the 
statistical and systematic uncertainties. Statistical errors
depend primarily on the signal-to-noise of the \hi\ spectrum, and are
obtained from the $rms$ noise of the linear fits to the edges of the
\hi\ profile. Systematic errors depend on the subjective choice of the
\hi\ signal boundaries \citep[see][]{gass1}, and are negligible for most of
the galaxies in our sample (see also Appendix~B).

Observed \hi\ widths are corrected for instrumental broadening and cosmological redshift as described in \citet{gass4}. \hi\ rotational velocities are then 
computed by deprojecting the corrected half-widths to edge-on view:
\begin{equation*}
    V_{\rm HI}= \frac{W^c_{50}}{2 \sin(i)},
\end{equation*}
\noindent
where $i$ is the inclination to the line of sight. Inclinations are 
determined from the SDSS \rband\ axis ratio (b/a) as:
\begin{equation*}
\cos(i) = \sqrt \frac{(b/a)^2 - q_0^2}{1-q_0^2},
\end{equation*}
\noindent
where $q_0$, the intrinsic axial ratio of a galaxy seen edge-on, is fixed at 0.2, appropriate for a thick disc \citep[e.g.,][]{guthrie92,giovanelli97a,weijmans14,tiley19}. We set the inclination to 90 degrees if $b/a<q_0$.
The catalogues of \hi\ parameters and spectra for our SAMI-\hi\ observations are presented in Appendix~A.

Lastly, we visually inspected the SDSS optical images of all the galaxies in our sample to determine which ones were affected by confusion within the Arecibo beam (full width half maximum size of \about 3.5 arcmin at 1.42 GHz). In these cases, 
two or more galaxies at very similar redshift lie within the radio beam, resulting in a blended signal from which we cannot reliably extract the \hi\ parameters of the target galaxy (in particular \hi\ flux and velocity width will be overestimated by
an unknown amount). In total, we flagged 25 galaxies (\about 8\% of the sample) for which confusion was certain, and identified these as systems with at least one late-type, similar size companion based on SDSS spectroscopy (i.e. with spectroscopic redshift available and within 0.002 of that of the target galaxy) within the beam. For our Arecibo observations, confusion is noted in Appendix~B.

\section{\Ha\ velocity field measurements from SAMI}\label{s_rcs}

The extraction of \Ha\ rotation curves from SAMI data and our procedure to measure \Ha\ rotational velocities are described in detail in \citet{tiley19}. Briefly, we start from the 2D line-of-sight emission line velocity maps obtained by fitting the LZIFU code \citep{lzifu} to the SAMI cubes (see \citealt{sami_dr3} for more details on the fitting procedure). Each SAMI velocity map is then centered on the peak of the continuum map, and the kinematic position angle is determined by using the maximum velocity gradient technique (i.e., by rotating the centred velocity map in 1-degree increments, and finding the value that maximizes the velocity gradient along
a 3 pixel-wide horizontal slit). About 10\% of these kinematic position angles were replaced after visual inspection. A rotation curve is then extracted from the velocity maps in a 3 pixel-wide slit along the major kinematic axis, by taking the weighted mean of the velocity values in pixel-wide steps within the slit.

We fit the SAMI \Ha\ rotation curves with the commonly used arctangent model \citep{courteau97}, which is left free to shift along the radial and velocity axes within 1.5 arcsec and 200 \kms, respectively. \Ha\ rotational velocities are measured 
from the best-fitting model at 1.3~\re\ (sometimes requiring extrapolation, see \S~\ref{s_vel}), where \re\ is the effective (half-light) radius convolved to the seeing of the velocity map. 
This radius corresponds to 2.2 disk scale lengths, where the rotation curve of a pure exponential disk would peak \citep{freeman70}. Uncertainties on these velocities, which we refer to as $v_{2.2,obs}$, are obtained by bootstrap re-sampling the observed rotation curve and refitting 100 times.
We also measured the maximum radial extent of the \Ha\ emission along the kinematic axis, \rmax, and the corresponding observed and model velocity at this radius.
Lastly, we correct the observed velocities for the effects of inclination and beam smearing using equation~6 in \citet{tiley19}:
\begin{equation*}
    V'_{\rm OPT} = \frac{\epsilon_{\rm R,PSF}}{\sin(i)} ~v_{2.2,obs},
\end{equation*}
\noindent
where $i$ is the inclination of the galaxy (see \S~\ref{s_obs}) and $\epsilon_{\rm R,PSF}$ is a beam smearing correction factor that depends on the ratio of the galaxy size to the width of the seeing point spread function (PSF), and the rotation speed of the galaxy. To be consistent with \hi\ measurements, we also correct optical rotational velocities for cosmological redshift:
\begin{equation*}
   V_{\rm OPT} = V'_{\rm OPT}/(1+z),
\end{equation*}
\noindent
where $z$ is the redshift of the galaxy measured from its \hi\ spectrum.

\begin{figure*}
\begin{center}
\includegraphics[width=17cm]{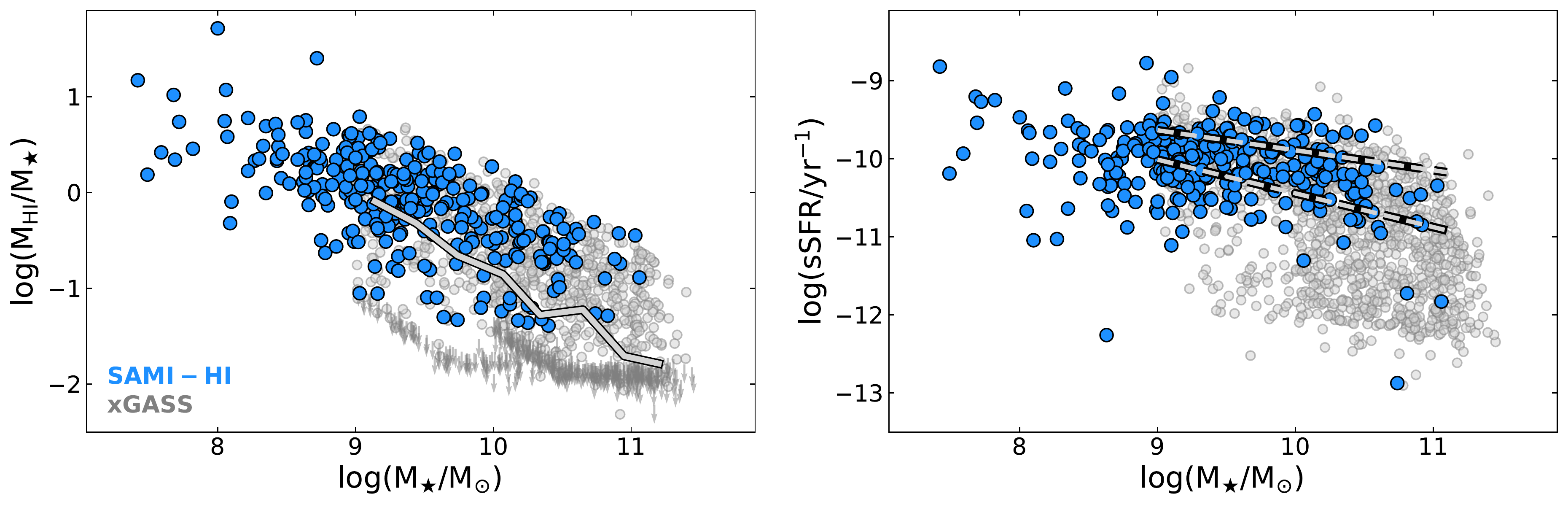}
\caption{{\it Left}: \hi\ gas fraction versus stellar mass for SAMI-\hi\ (blue) and, for comparison, xGASS (gray circles: \hi\ detections; gray downward arrows: upper limits). The solid line indicates xGASS median gas fractions, computed including upper limits of non-detections. {\it Right}: specific SFR as a function of stellar mass for SAMI-\hi\ (blue) and xGASS (gray). Dashed lines enclose the xGASS star-forming sequence and correspond to $\pm 1\sigma$ deviation from the average.
}
\label{fig2}
\end{center}
\end{figure*}

\section{Results}\label{s_res}

\subsection{The SAMI-\hi\ sample}\label{s_sample}

The sky distribution of the SAMI-\hi\ sample is shown on the top panels of Fig.~\ref{fig1}, with orange and gray circles indicating the subsets with new Arecibo data and ALFALFA detections, respectively. The galaxies are roughly evenly distributed among the three equatorial GAMA fields centered at right ascensions of 9 hr (left), 12 hr (center) and 14.5 hr (right). The bottom panel shows the distribution of \hi\ mass as a function of optical spectroscopic redshift for the same subsets. As can be seen, our targeted Arecibo observations detected galaxies with \hi\ masses \about 1 dex smaller than ALFALFA at similar redshift. Unfortunately the $z>0.06$ redshift interval was not accessible due to RFI.

In order to better characterise the properties of SAMI-\hi, we compare it with xGASS, a reference sample of \about 1200 galaxies with homogeneously measured \hi\ and multi-wavelength parameters. Briefly, galaxies in xGASS were selected by redshift ($0.01\leq z\leq 0.05$) and stellar mass ($9 \leq \log(M_\star/\rm M_\odot) \leq 11.5$) only, forcing an approximately flat distribution in \Mst\ (in each of the two volumes that make up the survey, i.e. $0.01\leq z\leq 0.02$ and $0.025\leq z\leq 0.05$) to increase statistics at the high \Mst\ end. The targets were observed with Arecibo until detected, or until a gas fraction limit of a few per cent in gas mass fraction (\Mhi/\Mst) was reached, making these the most sensitive \hi\ observations of a large, representative sample to date. 
Fig.~\ref{fig2} compares SAMI-\hi\ (blue circles) and xGASS (gray symbols) on the gas fraction-stellar mass and specific SFR vs. stellar mass planes. On the left panel, small downward arrows are xGASS non-detections plotted at their 5$\sigma$ upper limits and 
the gray line shows the xGASS medians (computed including the upper limits; see \citealt{xgass}). 
SAMI-\hi\ probes stellar masses almost 2 dex below xGASS, and does not reach quite the same high stellar masses as xGASS (this is because xGASS oversampled the high \Mst\ regime), but where the two samples overlap, there is a clear offset. Specifically, SAMI-\hi\ galaxies have systematically higher gas fractions 
than xGASS at the same stellar mass, and are for the most part star-forming systems with specific SFRs above $\rm 10^{-11}~yr^{-1}$ (this is by selection, see \S~\ref{s_sel}). For reference, the xGASS star-forming main sequence is indicated by the dashed lines on the right panel, which are at $\pm 1\sigma$ from the average relation (see \citealt{xgass} and \citealt{janowiecki20} for details).

\begin{figure}
\begin{center}
\includegraphics[width=8.5cm]{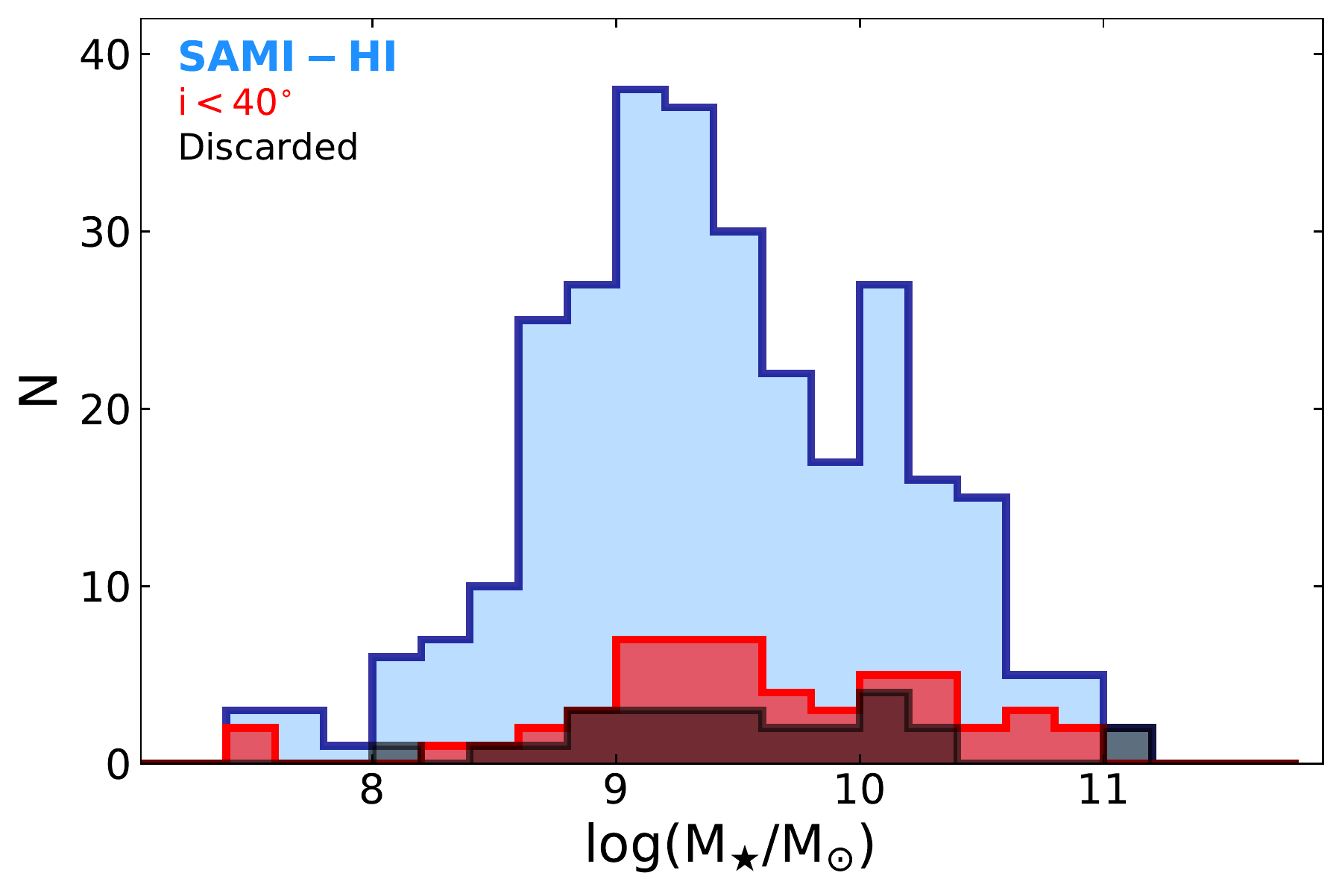}
\caption{Stellar mass histograms showing the selection cuts applied to the sample. Blue: SAMI-\hi\ sample (N=296). Dark gray: galaxies that were discarded because of invalid \Ha\ or \hi\ rotational velocities (N=27, see \S~\ref{s_sample}). Red: low-inclination systems (N=56).
}
\label{fig3}
\end{center}
\end{figure}

In what follows, we discard 27 galaxies from the SAMI-\hi\ sample: (a) 25 systems that are confused within the Arecibo beam (see \S~\ref{s_obs}); (b) one galaxy with incorrect \hi\ parameters (SAMI 298862/AGC 243504, for which ALFALFA measured only one of the two \hi\ peaks, severely underestimating its velocity width); and (c) one galaxy with a measured \Ha\ rotational velocity of 0 \kms\ (SAMI 278684). Discarding these systems does not introduce significant biases (see Fig.~\ref{fig3}). We refer to the resulting subset, which includes 269 galaxies, as the SAMI-\hi\ {\it kinematic} sample. As discussed in \S~\ref{s_tfr}, the {\it pruned} SAMI-\hi\ kinematic subset additionally excludes 56 galaxies with inclinations $i < 40$\deg\ and 12 systems with disturbed \Ha\ kinematics (2 of which also have $i < 40$\deg), thus including 203 galaxies.

\subsection{\Ha\ and \hi\ rotational velocity comparison}\label{s_vel}

\begin{figure*}
\begin{center}
\includegraphics[width=17.5cm]{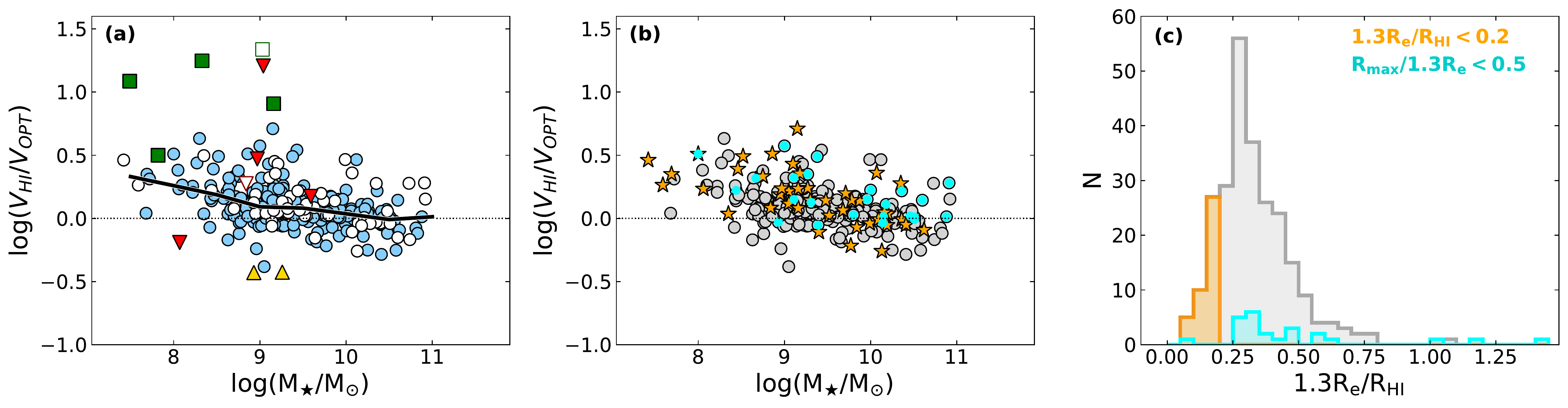}
\caption{{\bf (a)} Ratio of \hi-to-\Ha\ rotational velocities (independent of inclination) versus stellar mass for the SAMI-\hi\ kinematic sample. Empty symbols indicate galaxies with low inclinations ($i<40$\deg), and colored symbols are problematic systems identified by visual inspection (green squares: mergers/messy; red downward triangles: outflows/counter-rotating; yellow triangles: wrong GAMA radii). The thick line shows the medians for the full sample. {\bf (b)} Same as (a) but with problematic systems (squares and triangles) removed. Points here are colour-coded to highlight galaxies with \vopt\ measured at a radius that is a small fraction of the \hi\ size (i.e., 1.3 \re/\rhi $<0.2$; orange stars) and galaxies with the largest rotation curve extrapolations (i.e., \rmax/(1.3 \re) $< 0.5$, where \rmax\ is the extent of the rotation curve; cyan circles). {\bf (c)} Distributions of 1.3 \re/\rhi, i.e. the ratio of the radii where \vopt\ and \vhi\ are measured (the size of the \hi\ disk is estimated from the \hi\ mass-size relation, see \S~\ref{s_vel}) for the 3 subsets of galaxies shown in (b).
}
\label{fig_vratio}
\end{center}
\end{figure*}

Before presenting the TFRs for our sample in the next section, we compare the rotational velocities obtained from SAMI \Ha\ rotation curves (\vopt, measured at 1.3 \re) and from Arecibo \hi\ global profiles (\vhi, measured at 50\% of the peaks). Fig.~\ref{fig_vratio}a shows the ratio between the two, \vhi/\vopt, plotted as a function of stellar mass for the SAMI-\hi\ kinematic sample. While this velocity ratio does not depend on the galaxy's inclination to the line-of-sight, we indicate low-inclination systems ($i<40$\deg) with empty symbols, as these are typically excluded in TFR studies.
As expected from the fact that \hi\ measurements probe larger radii (and rotation curves are typically still rising at 1.3 \re), most galaxies have \hi\ rotational velocities that are larger than optical ones.
The \vhi/\vopt\ ratio shows a clear dependence on stellar mass, in the sense that the lower \Mst\ is, the more \vopt\ underestimates \vhi. 
This is quantified by the running medians (thick line); a linear fit to the gray points (not shown) has a correlation coefficient $r=-0.42$ and a scatter $\sigma=0.15$ dex. The reason for this trend with stellar mass is, again, due to a combination of rotation curve shapes and extent probed by the velocity tracers. When binned by galaxy luminosity and plotted as a function of normalised radius (e.g., galactocentric radius divided by scale length or optical radius), average rotation curve shapes vary smoothly, from slowly rising at low luminosity, to steeply rising in the central parts and then approximately flat in the brightest galaxies \citep[e.g.,][]{verheijen01,templates}. Thus, the 
ratio between rotational velocities measured in the outer parts and at 1.3 \re\ 
increases at low stellar masses, where rotation curves are typically rising and have not reached the flat part. This implies systematic differences in the slopes of TFRs based on \Ha\ and \hi\ measurements, as discussed in the next section.

It is interesting to note that there are galaxies with \hi\ velocity widths that are smaller than optical ones. While at high stellar masses these galaxies are within the scatter of the relation, at lower masses
($\log(M_\star/\rm M_\odot) \lesssim 9.5$) these are clearly outliers. 
We inspected the galaxies with $\log (V_{\rm HI}/V_{\rm OPT})<-0.1$ across the full \Mst\ range, and concluded that these tend to be the low SNR tail of the \hi-detected population (about 80\% are ALFALFA detections, half of which were detected with relatively low SNR of $6-8$). The strongest outliers, such as the two galaxies with \vhi/\vopt \about 0.5 at 
$\log(M_\star/\rm M_\odot) \sim 9$ (blue circles closest to the two yellow triangles), can be explained as having \vhi\ potentially underestimated. This illustrates some of the challenges involved in a detailed comparison between \Ha\ and \hi\ velocity widths, even for reliable \hi\ detections. At the high stellar mass end, a few outliers can be explained by overestimated optical velocities, for instance due to large extrapolations in galaxies where the rotation curve traces a strong bar.

Fig.~\ref{fig_vratio}a also shows a few strong outliers above the relation.
We visually inspected the SDSS images and SAMI velocity fields of several galaxies that were outliers of this relation and/or the TFRs presented below, and identified two systems with GAMA radii clearly wrong (yellow triangles in the figure) and nine with highly disturbed \Ha\ velocity fields, due to mergers
(green boxes) and presence of outflows or counter-rotation (downward red triangles). These galaxies are shown in Appendix~C and should not be plotted on the TFR,
as \vopt\ is not a good measurement of rotational velocity.

The sample obtained after removing the systems marked with colored squares or triangles in Fig.~\ref{fig_vratio}a is shown in panel (b), without distinction by inclination. One might expect that the scatter of this relation could be in part related to the ratio of the radii where \vopt\ and \vhi\ are measured, which could vary significantly from galaxy to galaxy. In order to test this hypothesis, 
we take advantage of the very tight \hi\ mass-size relation by \citet{jing16} to estimate the size of the \hi\ disk from the global \hi\ mass:
\begin{equation*}
    \log D_{\rm HI} = (0.506 \pm 0.003) \log M_{\rm HI} - (3.293 \pm 0.009),
\end{equation*}
\noindent
where $D_{\rm HI}$ is the diameter in kpc of the \hi\ disk measured at a surface density 
$\Sigma_{\rm HI}=1~\rm M_\odot ~pc^{-2}$. This relation holds over \about 5 dex in \hi\ mass and has a scatter of only \about 0.06 dex \citep[see also][]{stevens19}.

The distribution of the 1.3 \re/\rhi\ ratio (where $R_{\rm HI}=0.5~D_{\rm HI}$) for this sample is presented in Fig.~\ref{fig_vratio}c, and indeed shows a significant variation, with four galaxies having estimated \hi\ radii that are shorter than 1.3 \re. Galaxies with 1.3 \re/\rhi\ ratio smaller than 0.2 (16\% of the sample, orange histogram) do not appear to scatter more than the rest of the sample in panel (b), and are rather evenly distributed in stellar mass (orange stars).

Another factor that could contribute to the scatter of the velocity ratios is the extent of the \Ha\ emission,
\rmax, as rotation curves that do 
not reach 1.3 \re\ require an extrapolation to estimate \vopt. Galaxies with the largest rotation curve extrapolations (i.e., for which \rmax/(1.3 \re) $< 0.5$) are marked in cyan in panels (b) and (c). Comparing panels (a) and (b) it is interesting to note that, with the exception of a couple of galaxies, all the outliers above the relation in panel (b) are marked in cyan (largest rotation curve extrapolations) and/or orange (smallest 1.3 \re/\rhi\ ratios) and/or have low inclinations (empty circles in (a)).

As shown in Fig.~\ref{fig_vratio}, a small fraction of galaxies required significant extrapolation of their \Ha\ rotation curves to measure \vopt\ (9\% of the sample, cyan symbols and histogram). For 49\% of SAMI-\hi, rotation curves extend beyond 1.3 \re, therefore one might wonder if it would be preferable to measure \vopt\ at \rmax\ instead. While measuring rotational velocities at larger radii (where rotation curves are flatter) is better, it is equally important to determine these quantities in a consistent way (in this case, at the same relative position along the major axis of the disk) for the full sample, in order to avoid systematic effects that vary with rotation curve shape and extent \citep[e.g.,][]{widths}.

\begin{figure*}
\begin{center}
\includegraphics[width=17.5cm]{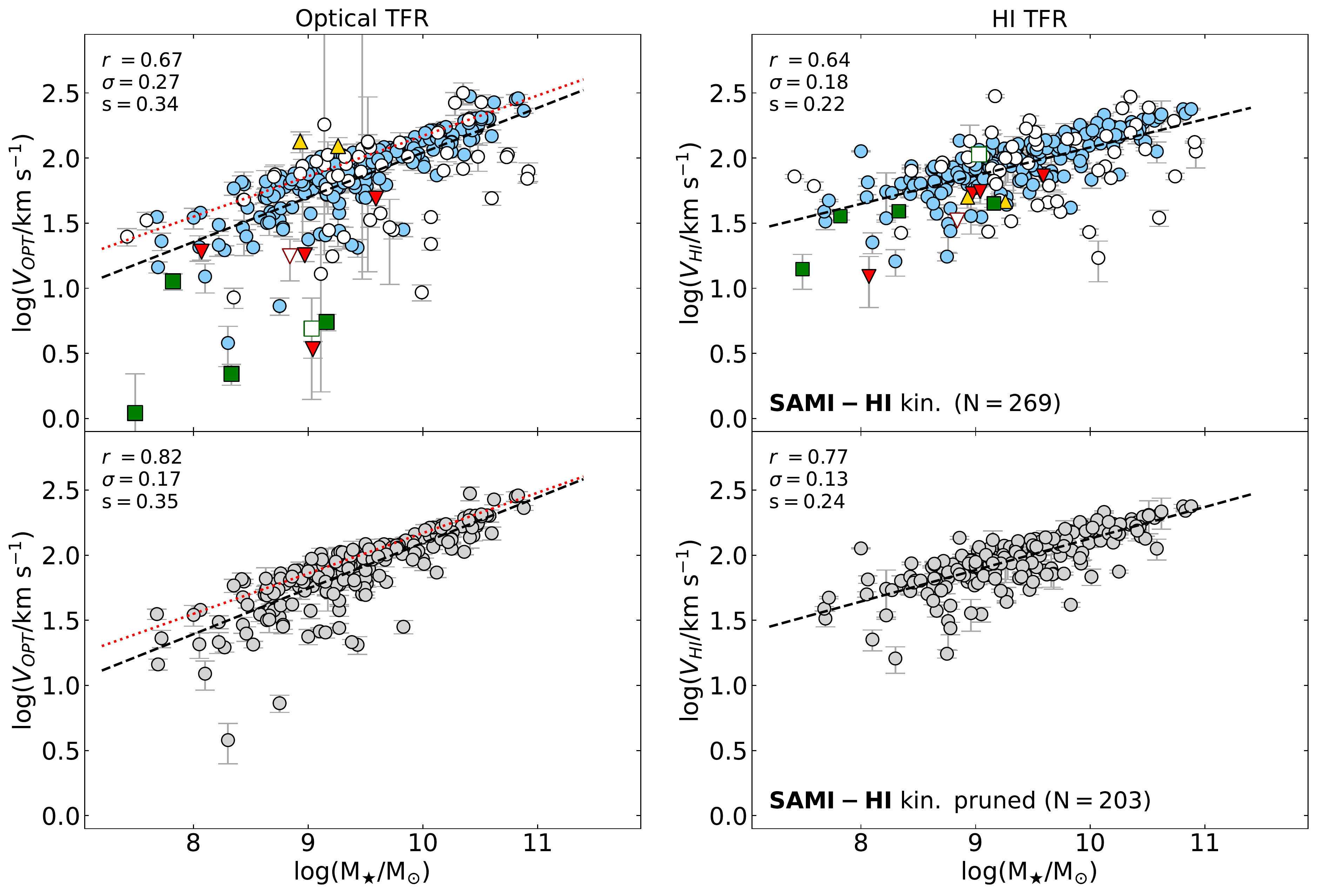}
\caption{{\it Top:} optical (left) and \hi\ (right) stellar mass TFRs for the  SAMI-\hi\ kinematic sample (same symbols as in Fig.~\ref{fig_vratio}a). Dashed lines are direct linear fits, with correlation coefficient, scatter and slope indicated on the top-left corners (see also Table~\ref{t_fits}). {\it Bottom:} TFRs for the kinematic, pruned sample, obtained by removing galaxies with inclination $i<40$\deg\ and problematic systems (identified by squares and triangles in the top panels and in Fig.~\ref{fig_vratio}a). Red dotted lines on the left panels show the fit to a larger sample of SAMI galaxies from \citet{bloom17}.
}
\label{fig_tfr}
\end{center}
\end{figure*}

\subsection{Stellar and baryonic mass Tully-Fisher relations}\label{s_tfr}

As typically done in optical studies, we present here TFRs by plotting rotational velocities as a function of stellar mass.
This is in contrast to traditional TFR \hi\ studies, where luminosity or mass is treated as the dependent variable for distance indicator applications, and inverse fits are used to minimise the effect of Malmquist bias \citep[e.g.,][]{teerikorpi84,teerikorpi97,giovanelli97b}.
Optical and \hi\ stellar mass TFRs based on the SAMI-\hi\ kinematic sample are compared in the top panels of Fig.~\ref{fig_tfr}, with the same symbols as in Fig.~\ref{fig_vratio}a. Error bars on the optical and \hi\ rotational velocities are computed as explained in \S~\ref{s_rcs} and \S~\ref{s_obs}, respectively, and do not include errors on the inclinations, which are dominated by the unknown uncertainty on the assumed intrinsic disc axial ratio.
Dashed lines show direct fits to all the data (not including errors), with correlation coefficient, scatter and slope noted on the top-left corner of each panel (see also Table~\ref{t_fits}). In the top-left panel, a red dotted line indicates the fiducial fit to the TFR obtained by \citet{bloom17} for a sample of \about 700 SAMI galaxies, after excluding the low-mass (high kinematic asymmetry) outliers (their equation~4):
$\log(V_{\rm rot}/$\kms) $= (0.31\pm 0.01) \log(M_\star/$\Msun) $-0.93\pm 0.10$.
If we excluded the low-mass outliers below the TFR from our fit as they did, we would obtain a result more consistent with \citet{bloom17}.
Unsurprisingly, both \Ha\ and \hi\ TFRs have significant scatter (\about 0.3 and 0.2 dex, respectively), as our sample was not designed to obtain the tightest TFR, as done for example for distance indicator applications. Galaxies that are typically excluded from TFR studies include low-inclination systems (i.e., $i<40$\deg, indicated by empty symbols in the top panels), for which the deprojection to line-of-sight rotational velocity is particularly uncertain. Galaxies with complex \Ha\ kinematics, for which \vopt\ is not a reliable measure of rotational velocity (green and red symbols), are preferentially found at low stellar masses and are amongst the strongest outliers of the optical TFR, whereas they do not deviate significantly on the \hi\ TFR.

The bottom two panels of Fig.~\ref{fig_tfr} show the pruned versions of the two TFRs, after removing galaxies with $i<40$\deg\ (empty symbols) and/or with disturbed \Ha\ kinematics (colored squares ans triangles). The subset of galaxies with low inclinations includes 13 galaxies with inclinations that were largely overestimated, typically due to the presence of a strong bar; these were identified by visually inspecting the SDSS images of all the remaining outliers after the first pruning, and are all galaxies with \Mst $>10^9$ \Msun\ that would have been excluded by our $i<40$\deg\ cut (see Appendix~C for more details). 

\begin{table}
\centering
\caption{Fits to the stellar and baryonic TFRs presented in Fig.~\ref{fig_tfr} (top four rows) and Fig.~\ref{fig_btf} (bottom two rows).
We fit the linear relation $\log(V/$\kms) $= s \log(M/$\Msun) + $q$, where $V$ is \vopt\ or \vhi\ for the optical or \hi\ relations, and $M$ is \Mst\ or $M_{\rm bar}$ for stellar or baryonic TFRs, respectively. The scatter $\sigma$ in $\log(V/$\kms) and the size of the sample are listed in the last two columns.}
\label{t_fits}
\begin{tabular}{lcccc}
\hline
Relation          & $s$             &  $q$              & $\sigma$ &   N \\
\hline
OPT TFR           & 0.34 $\pm$ 0.03 & $-$1.39 $\pm$ 0.27 &  0.27 &  269 \\
\hi\ TFR          & 0.22 $\pm$ 0.02 & $-$0.09 $\pm$ 0.17 &  0.18 &  269 \\
OPT TFR, pruned   & 0.35 $\pm$ 0.02 & $-$1.41 $\pm$ 0.18 &  0.17 &  203 \\
\hi\ TFR, pruned  & 0.24 $\pm$ 0.01 & $-$0.29 $\pm$ 0.14 &  0.13 &  203 \\
OPT BTFR, pruned  & 0.44 $\pm$ 0.02 & $-$2.43 $\pm$ 0.24 &  0.17 &  203 \\
\hi\ BTFR, pruned & 0.32 $\pm$ 0.02 & $-$1.17 $\pm$ 0.17 &  0.13 &  203 \\
\hline
\end{tabular}
\end{table}

While both \Ha\ and \hi\ relations have tightened (by \about 0.1 and 0.05 dex, respectively) compared to the versions in the top panels, the \Ha\ TFR (left) shows a number of outliers at low stellar mass (\Mst $\lesssim 10^{9.5}$ \Msun) that are not seen in the \hi\ TFR (right). In \hi\ TFR studies it is well known that dwarf galaxies (\Mst $\lesssim 10^9$ \Msun) systematically deviate from the main relation defined by larger systems, in the sense that, at fixed rotational velocity (i.e., dynamical mass), their stellar masses are smaller than expected. This is interpreted as due to the fact that a larger fraction of the baryonic mass (i.e., the sum of stellar and cold gas mass) in dwarf galaxies is in the form of cold gas, and indeed a tight and linear TFR is recovered when baryonic mass is plotted instead of stellar mass \citep{mcgaugh00}. Interestingly, the low \Mst\ outliers of the \Ha\ TFR scatter in the {\it opposite} direction: at fixed rotational velocity, stellar masses are higher than expected based on the fit to the larger systems.

\begin{figure}
\begin{center}
\includegraphics[width=8.5cm]{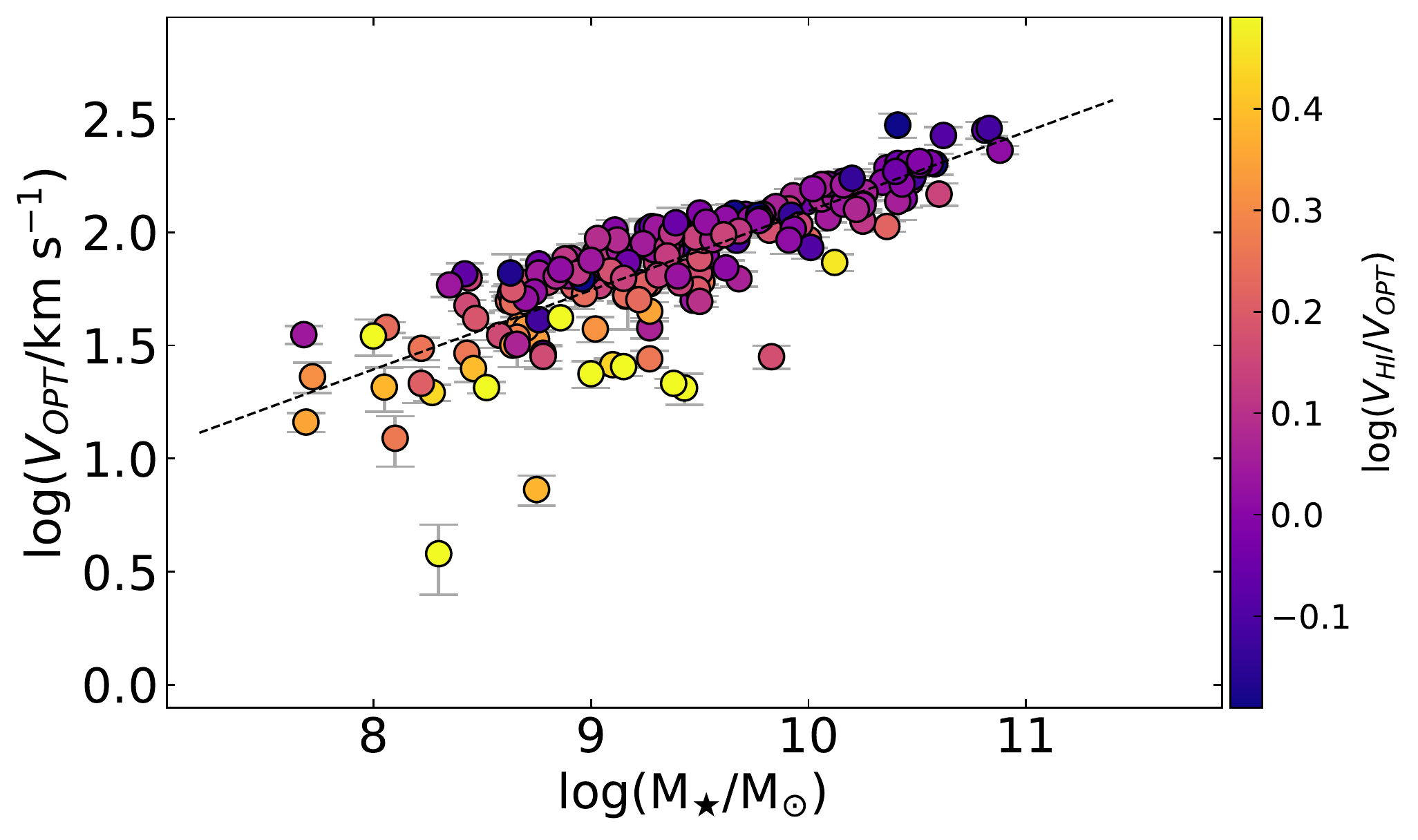}
\caption{Optical stellar mass TFR of the SAMI-\hi\ kinematic, pruned sample and direct linear fit, reproduced from the bottom left panel of Fig.~\ref{fig_tfr}, with points color-coded by the logarithmic ratio of \hi-to-\Ha\ rotational velocities. 
}
\label{fig_tfr_color}
\end{center}
\end{figure}

As mentioned in the introduction, the increased scatter at low stellar mass of the optical TFR has been attributed to an increased importance of the velocity dispersion $\sigma$ in those systems (i.e., a decrease of \vs; \citealt{simons15,bloom17}). In other words, a non-negligible part of the dynamical support in low-mass galaxies comes from random motions, not captured by \vopt. While this might be the case, other explanations are also possible. 
IFS samples such as SAMI cover only the inner parts of the galaxies, and velocities estimated at a fixed radius (generally within 2 effective radii) might systematically underestimate the rotational velocity of galaxies compared to global \hi\ measurements, which probe larger radii. This is especially true at low stellar masses, where rotation curves are typically rising and have not reached the flat part \citep[e.g.,][]{verheijen01,templates}. This is indeed demonstrated in Fig.~\ref{fig_tfr_color}, where we color-coded the SAMI-\hi\ (pruned) optical TFR by the ratio of \hi\ and \Ha\ rotational velocities -- the low-mass outliers of the relation are the galaxies for which \vopt\ underestimates \vhi\ the most (although the opposite is not true -- not all the systems with large \vhi/\vopt\ ratios are outliers). 

Interestingly, while long-slit spectroscopy observations do not have the same aperture limitation as IFS, they too are likely to underestimate the rotational velocity, as a result of any alignment error between the photometric and the kinematic position angles (the latter being unknown a-priori).
Either way, optical spectroscopy of faint galaxies is challenging and more prone to measurement errors. This is not necessarily the case with \hi\ observations -- low mass galaxies are more gas rich (per unit of stellar mass, see Fig.~\ref{fig2}) and have smaller velocity widths, making them easier to detect above the noise (compared to a galaxy with a the same \hi\ mass and larger stellar mass). We discuss below whether the low-mass outliers of the optical TFR have a physical explanation or are the result of measurements issues (errors or aperture effects).

Fig.~\ref{fig_tfr} also shows that the slopes of the two TFRs are different, with the \hi\ relation being shallower (in this optical TFR representation, where \Mst\ is the independent variable).
This is caused by the combination of rotation curve shapes and rotational velocity definition, as discussed in the previous section.

\begin{figure*}
\begin{center}
\includegraphics[width=17.5cm]{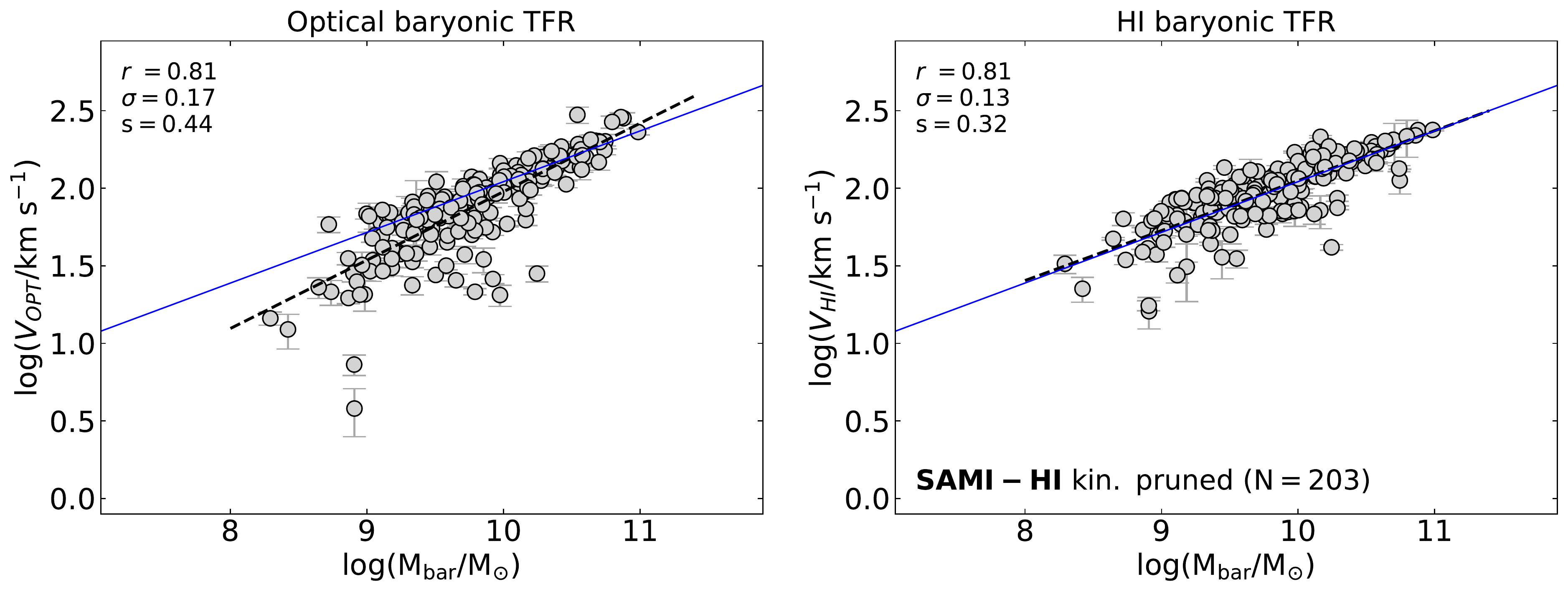}
\caption{Optical (left) and \hi\ (right) baryonic mass TFRs for the SAMI-\hi\ kinematic, pruned sample. Dashed lines in each panel are direct linear fits, with correlation coefficient, scatter and slope indicated on the top-left corners (see also Table~\ref{t_fits}). Blue solid lines indicate the relation from \citet{lelli19}.
}
\label{fig_btf}
\end{center}
\end{figure*}

For completeness, we show in Fig.~\ref{fig_btf} the \Ha\ and \hi\ baryonic TFRs for the SAMI-\hi\ kinematic, pruned sample. Baryonic masses are estimated as $M_{\rm bar}= M_\star\ + 1.36 M_{\rm HI}$, where the factor 1.36 takes the contribution of Helium into account. Overlaid on both plots is the relation obtained by \citet[][see their Table~1]{lelli19} using optical rotational velocities measured at 2.2 disk scale lengths:
$\log(M_{bar}/$\Msun$)= (3.06 \pm 0.08) \log(V_{2.2}/$\kms) $+3.75 \pm 0.17$, which is indistinguishable from our own fit to the \hi\ relation. Compared to the stellar TFRs, the baryonic TFRs are steeper but the scatters are unchanged. The increased slopes are simply a consequence of the fact that \hi\ gas fractions are higher at lower stellar masses \citep[e.g.,][]{xgass,SC22_araa}, thus lower \Mst\ points shift rightwards by a larger amount. Most importantly, the optical baryonic TFR still has low \vopt\ outliers that are not seen in the \hi\ relation.

\section{Discussion and conclusions}\label{s_concl}

In this paper we presented SAMI-\hi, an Arecibo program that measured global \hi\ parameters for 296 galaxies selected from the SAMI Galaxy Survey. In addition to releasing the new \hi\ data, we used this sample
to illustrate the power, but also the challenges, of combining 3D optical data from large multiplexed IFS surveys with 21 cm single-dish observations to better understand dynamical scaling relations in the local Universe. While comparing IFS with spatially-resolved \hi\ data would be a more insightful exercise, the vast majority of the galaxies that will be detected by the next-generation \hi-line surveys with the SKA and its pathfinders will be spatially unresolved or only marginally resolved.
\citep[e.g.,][]{SC22_araa}. Thus, our analysis is illustrative of what will be typically feasible once large \hi\ programs 
like WALLABY, DINGO and LADUMA enter full survey mode.

Our comparison of \hi\ and \Ha\ rotational velocities builds on previous work \citep[e.g.,][]{courteau97,widths,lelli19} and confirms how these estimates strongly depend on the tracer used and on the radial position at which the velocities are measured. Systematic differences between \vopt\ measured at 1.3 effective radii from IFS observations and \vhi\ (which is closer to the maximum rotational velocity of galaxies) automatically lead to differences in slopes (and zero points) of dynamical scaling relations such as the TFR. While this has been noted in the past, for instance in comparisons between dynamical scaling relations obtained from \hi\ widths and optical long-slit spectroscopy, our analysis reframes the issue in the context of large IFS datasets and upcoming SKA-era \hi\ surveys and shows how a quantitative comparison of dynamic scaling relations obtained from these samples is far from trivial. Naturally, this applies not only to TFR analyses in the local Universe, but also to studies of the shape and cosmic evolution of other dynamical scaling relations (such as the mass-specific angular momentum relation; e.g., \citealp{fall83,romanowsky12,luca16a,tiley21,hardwick22}) and to comparisons with numerical simulations. Taking also into account how notoriously challenging it is to determine homogeneous velocity widths as a function of redshift even from the same data \citep{tiley19}, it is clear that extreme care must be taken in order to interpret offsets and differences in dynamical scaling relations based on different tracers and/or data sets. Therefore, when it comes to evolution of scaling relations with cosmic time, our results suggest that we should avoid using different tracers for different populations of galaxies, and primarily focus on homogenising 
observations of the same tracer (either \hi, \Ha\ or other emission lines) across redshift, if possible.

On a more positive note, the analysis of our SAMI-\hi\ sample with both IFS and \hi\ data has provided us with some additional insights into the increase of scatter of the \Ha\ TFR at low stellar masses claimed in previous works \citep{simons15,bloom17}. As shown in Fig.~\ref{fig_tfr}, when \Ha\ rotational velocities are measured at 1.3 \re, the optical TFR has larger scatter than the \hi\ one, and this difference is more prominent at lower stellar masses (i.e., \Mst $\lesssim 10^{9.5}$~\Msun), although this scatter increase seems less dramatic than what is reported by other groups.
Most importantly, our inspection of the SAMI 2D velocity maps revealed that most of the significant low-mass outliers of the \Ha\ TFR are not regular rotators but kinematically disturbed systems, either merger remnants or tidally perturbed (see Fig.~\ref{fig_outliers}).   
Interestingly, from optical imaging these galaxies appear to be unperturbed dwarf irregulars, and thus were not flagged as potentially confused within the radio beam (and excluded from the analysis). 

Our findings support the conclusions of \citet{simons15} and \citet{bloom17}, who pinned
the increased scatter of the SAMI TFR at low stellar masses on the outliers with 
higher kinematic asymmetries compared to galaxies on the main relation. However,
our view is that these systems should not be included in a TFR sample, because \Ha\ widths measured from highly disturbed velocity fields should not be blindly interpreted as reliable estimates of rotational velocities.
This becomes even more challenging with \Ha\ velocity widths measured from long-slit spectroscopy, as the loss of spatial information compared to IFS data makes it 
harder to judge the degree of disturbance in the velocity field.
This issue, combined with the fact that dwarf galaxies have intrinsically small rotational velocities at 1.3~\re\ (comparable or even smaller than the \Ha\ gas velocity dispersion), raises the question of how meaningful is to look at the stellar TFR at low stellar masses, where rotational velocity estimates are highly uncertain and encompass a tiny fraction of the total dynamical mass of the galaxies. 
As shown by various works (e.g., \citealp{kassin07,luca14, barat19,barat20,aquino18,lelli19}), in this regime it is preferable to investigate 
dynamical scaling relations expressed in terms of circular velocity or S$_{05}$ parameter (which includes contributions from both rotational velocity and velocity dispersion), for which no increase in scatter at low masses has been confirmed so far. 

In conclusion, below \Mst \about 10$^{9}$~\Msun\ dynamical scaling relations are best investigated with \hi\ velocity widths. This is especially true for the TFR because, at those stellar masses, galaxies are \hi-dominated and the baryonic (rather than stellar) TFR is a more physical relation \citep{mcgaugh00}. Optical IFS observations are catching up but, in order to minimise limitations of current data sets, it is paramount to trace the rotation curves beyond 1.3~\re\ and, ideally, with spectral resolutions significantly higher those of surveys like SAMI and closer to the typical values of \hi\ observations.

\section*{Acknowledgments}

We thank the anonymous referee for constructive comments.
BC thanks Katinka Ger\'eb for help with the observations and Danail Obreschkow for useful comments. 
LC is the recipient of an Australian Research Council Future Fellowship (FT180100066) funded by the Australian Government.
LC and ABW acknowledge support from the Australian Research
Council Discovery Project funding scheme (DP210100337).
ALT acknowledges support from a Forrest Research Foundation Fellowship.
JJB acknowledges support of an Australian Research Council Future Fellowship (FT180100231). 
FDE acknowledges funding through the ERC Advanced grant 695671 ``QUENCH", the H2020 ERC Consolidator Grant 683184 and support by the Science and Technology Facilities Council (STFC). 
JvdS acknowledges support of an Australian Research Council Discovery Early Career Research Award (project number DE200100461) funded by the Australian Government.
JBH is supported by an ARC Laureate Fellowship FL140100278. The SAMI instrument was funded by Bland-Hawthorn's former Federation Fellowship FF0776384, an ARC LIEF grant LE130100198 (PI Bland-Hawthorn) and funding from the Anglo-Australian Observatory. 
SMS acknowledges funding from the Australian Research Council (DE220100003). 
DJP is supported through the South African Research Chairs Initiative of the Department of Science and Technology and National Research Foundation. DJP, NMP, DC, and MCH acknowledge partial support from NSF CAREER grant AST-1149491.
NSF grants AST-1211005 and AST-1637339 are acknowledged for their support of members of the Undergraduate ALFALFA Team to observe at Arecibo Observatory.
Parts of this research were conducted by the Australian Research Council Centre of Excellence for All Sky Astrophysics in 3 Dimensions (ASTRO 3D), through project number CE170100013.

This research has made use of the NASA/IPAC Extragalactic Database
(NED) which is operated by the Jet Propulsion Laboratory, California
Institute of Technology, under contract with the National Aeronautics
and Space Administration.

The Arecibo Observatory is operated by SRI International under a
cooperative agreement with the National Science Foundation
(AST-1100968), and in alliance with Ana G. M{\'e}ndez-Universidad
Metropolitana, and the Universities Space Research Association.

The SAMI Galaxy Survey is based on observations made at the Anglo-Australian Telescope. 
The SAMI spectrograph was developed jointly by the University of Sydney and the Australian 
Astronomical Observatory. The SAMI input catalogue is based on data taken from the 
Sloan Digital Sky Survey, the GAMA Survey and the VST ATLAS Survey. The SAMI Galaxy Survey 
is funded by the Australian Research Council Centre of Excellence for All-sky Astrophysics 
(CAASTRO), through project number CE110001020, and other participating institutions. 
The SAMI Galaxy Survey website is http://sami-survey.org/.

GAMA is a joint European-Australasian project based around a spectroscopic campaign using 
the Anglo-Australian Telescope. The GAMA input catalogue is based on data taken from the 
Sloan Digital Sky Survey and the UKIRT Infrared Deep Sky Survey. Complementary imaging of 
the GAMA regions is being obtained by a number of independent survey programmes including 
GALEX MIS, VST KiDS, VISTA VIKING, WISE, Herschel-ATLAS, GMRT and ASKAP providing UV to 
radio coverage. GAMA is funded by the STFC (UK), the ARC (Australia), the AAO, and the 
participating institutions. The GAMA website is http://www.gama-survey.org/.

\section*{Data availability}
The SAMI data cubes and value-added products used in this paper are available from Astronomical Optics’ Data Central service at \url{https://datacentral.org.au/}.
\hi\ and optical parameters for the SAMI-\hi\ sample will also be available on Data Central, along with digital spectra of our \hi\ observations, once the paper is published on MNRAS.




\bibliography{sami_hi_final}



\appendix

\section*{Appendix A: Data Release}

\setcounter{table}{0}
\renewcommand{\thetable}{A\arabic{table}} 

\setcounter{figure}{0}
\renewcommand{\thefigure}{A\arabic{figure}} 

We present here SDSS postage stamp images, \hi-line spectra and 
catalog of \hi\ parameters for the 153 SAMI-\hi\ galaxies observed by our Arecibo programs. 
The content of the table is described below; notes on individual objects
(marked with an asterisk in the second-last column of Table~\ref{t_det}) are reported in Appendix~B.\\

\noindent
{\bf \hi\ source catalog.}\\
This data release includes 153 detections. The measured \hi\ parameters for the detected galaxies are listed in Table~\ref{t_det}, ordered by increasing right ascension:\\

\bct
Cols. 1 and 2: SAMI and SDSS identifiers. \\

\bct
Col. 3: GAMA spectroscopic redshift, $z_{\rm spec}$. \\

\bct
Col. 4: on-source integration time of the Arecibo
observation, $T_{\rm on}$, in minutes. This number refers to
{\it on scans} that were actually combined, and does not account for
possible losses due to RFI excision (usually negligible). \\

\bct
Col. 5: velocity resolution of the final, smoothed spectrum in \kms. 
In general, lower signal-to-noise detections require more smoothing in order 
to better identify the edges and peaks of the \hi\ profiles, needed to measure
the \hi\ parameters.\\

\bct
Col. 6: redshift, $z$, measured from the \hi\ spectrum.
The error on the corresponding heliocentric velocity, $cz$, 
is half the error on the width, tabulated in the following column.\\

\bct
Col. 7: observed velocity width of the source line profile
in \kms, \whi, measured at the 50\% level of each peak
(see \S~\ref{s_obs}).
The error on the width is the sum in quadrature of the 
statistical and systematic uncertainties in \kms. 
Systematic errors depend on the subjective choice of the
\hi\ signal boundaries
and are negligible for most of the galaxies in our sample (see also Appendix~B).\\

\bct
Col. 8: velocity width corrected for instrumental broadening
and cosmological redshift only, $W_{50}^c$, in \kms\ (see
\citealt{gass_dr2} for details). No inclination or turbulent motion
corrections are applied.\\

\bct
Col. 9: integrated \hi-line flux density in Jy \kms, $F_{\rm HI} \equiv \int S~dv$, measured on the 
smoothed and baseline-subtracted spectrum (observed velocity frame). The reported uncertainty 
is the sum in quadrature of the statistical and systematic errors (see col. 7).
The statistical errors are calculated according to equation 2 of S05
(which includes the contribution from uncertainties in the baseline fit).\\

\bct
Col. 10: rms noise of the observation in mJy, measured on the
signal- and RFI-free portion of the smoothed spectrum.\\

\bct
Col. 11: signal-to-noise ratio of the \hi\ spectrum, S/N,
estimated following \citet{saintonge07} and adapted to the velocity
resolution of the spectrum. 
This is the definition of S/N adopted by ALFALFA, which accounts for the
fact that for the same peak flux a broader spectrum has more signal.\\

\bct
Col. 12: base-10 logarithm of the \hi\ mass, \Mhi, in solar
units, computed via: 
\begin{equation}
    \frac{M_{\rm HI}}{\rm M_{\odot}} = \frac{2.356\times 10^5}{(1+z)^2}
    \left[ \frac{d_{\rm L}(z)}{\rm Mpc}\right]^2
    \left(\frac{\int S~dv}{\rm Jy~km~s^{-1}} \right)
\label{eq_MHI}
\end{equation}
\noindent
where $d_{\rm L}(z)$ is the luminosity distance to the galaxy at
redshift $z$ as measured from the \hi\ spectrum in the observed velocity frame. \\

\bct
Col. 13: base-10 logarithm of the \hi\ mass fraction, \Mhi/\Mst.\\

\bct
Col. 14: quality flag, Q (1=good, 2=marginal and 5=confused). 
An asterisk indicates the presence of a note for the source in Appendix~B.
Code 1 indicates reliable detections, with a S/N ratio of order of
6.5 or higher. Marginal detections have lower S/N {\bc (between 5 and 6.5)}, thus more uncertain
\hi\ parameters, but are still secure detections, with \hi\ redshift
consistent with the SDSS one. We flag galaxies as ``confused'' when
most of the \hi\ emission is believed to originate from another source
within the Arecibo beam. For some of the galaxies, the presence of
small companions within the beam might contaminate (but is unlikely to
dominate) the \hi\ signal -- this is just noted in Appendix~B.\\

\bct
Col. 15: Arecibo project code.\\

\begin{figure*}
\begin{center}
\includegraphics[width=15cm]{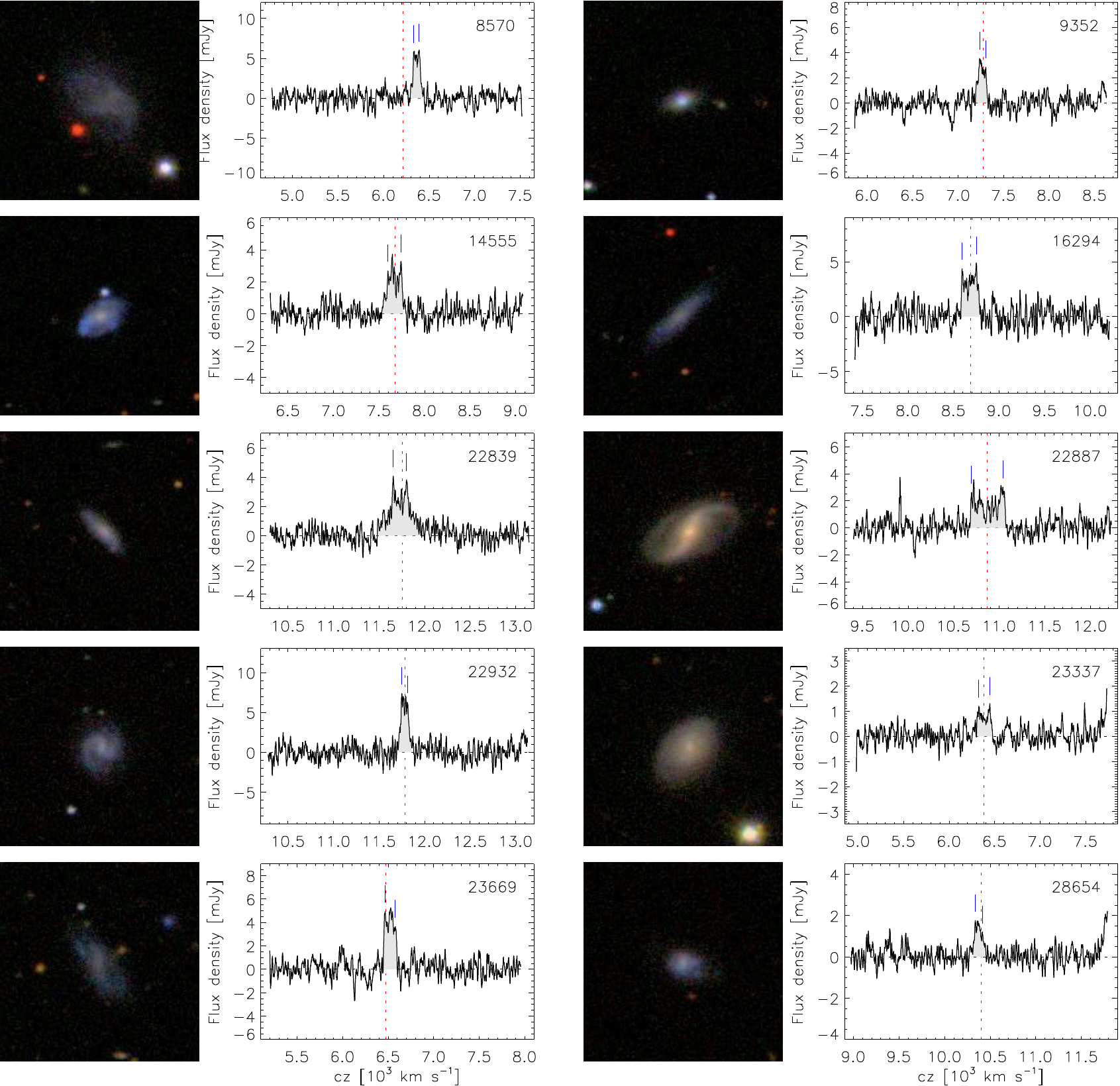}
\caption{SDSS postage stamp images (1 arcmin square) and
\hi-line profiles of SAMI-\hi\ galaxies, ordered by increasing SAMI number (indicated in each spectrum). 
The \hi\ spectra are calibrated, smoothed and baseline-subtracted. A dotted line and two
dashes indicate the heliocentric velocity corresponding to the GAMA optical
redshift and the two peaks used for width measurement, respectively. 
This is a sample of the complete figure, which is available in the online version of the article.}
\label{spectra}
\end{center}
\end{figure*}

\noindent
{\bf SDSS postage stamps and \hi\ spectra.}\\
Figure~\ref{spectra} shows SDSS images and Arecibo \hi\ spectra for a the first 10
galaxies included in this data release. The galaxies in each figure are ordered by 
increasing SAMI number, indicated on the top right corner of each spectrum.
The SDSS images show a 1 arcmin square field, \ie\ only the central part of the region 
sampled by the Arecibo beam (the half power full width of the beam is \about 3.5\arcmin\ at the
frequencies of our observations). Therefore, companions that might be detected in our 
spectra typically are not visible in the postage stamps, but they are noted in Appendix~B.
The \hi\ spectra are always displayed over a 3000 \kms\ velocity interval, which includes 
the full 12.5 MHz bandwidth adopted for our observations. The \hi-line profiles are calibrated, 
smoothed (to a velocity resolution between 5 and 15 \kms, as listed in Table~\ref{t_det}), and
baseline-subtracted. A red, dotted line indicates the heliocentric velocity corresponding to 
the optical spectroscopic redshift from GAMA. The shaded area and two vertical
dashes show the part of the profile that was integrated to measure the \hi\ flux and the peaks 
used for width measurement, respectively.

\onecolumn
\begin{table*}
\scriptsize
\centering
\caption{\hi\ Properties of SAMI-HI detections}
\label{t_det}
\begin{tabular}{ccccccccccrcclc}
\hline\hline
      &         &               & $T_{\rm on}$ & $\Delta v$ &     & \whi  & \whi$^c$&  $F$      &  rms & &log \Mhi  &   & & \\
SAMI  & SDSS ID & $z_{\rm spec}$ & (min)       &  (\kms)    & $z$ &  (\kms)& (\kms) & (Jy \kms) & (mJy)& S/N  & (\Msun) & log \Mhi/\Mst & Q & Proj\\
(1)  & (2)  & (3)  & (4)  & (5)  & (6)  & (7)  & (8)  & (9)  &  (10) & (11) & (12) & (13) & (14) & (15)\\
\hline
375402 & J083606.39+011823.0 & 0.0129 &  10 &  12 & 0.013166 & 112$\pm$   7 & 104 &  0.77$\pm$  0.05 &  0.53 &  28.1 &  8.76 &  0.33 &  1*  & a2934 \\
322910 & J083734.87+013426.0 & 0.0309 &  10 &   7 & 0.030971 &  51$\pm$   2 &  46 &  0.38$\pm$  0.03 &  0.73 &  19.1 &  9.19 & -0.55 &  1   & a2934 \\
300477 & J083849.62+010715.6 & 0.0291 &  15 &  13 & 0.029127 & 172$\pm$   2 & 161 &  0.39$\pm$  0.05 &  0.49 &  12.2 &  9.15 & -0.10 &  1   & a2934 \\
375531 & J083850.61+012418.0 & 0.0290 &   9 &  13 & 0.029117 & 148$\pm$  18 & 137 &  2.57$\pm$  0.07 &  0.73 &  58.2 &  9.97 &  0.52 &  5*  & a3116 \\
422289 & J084106.54+023436.0 & 0.0318 &  30 &  13 & 0.031422 & 204$\pm$   8 & 192 &  0.31$\pm$  0.04 &  0.35 &  12.4 &  9.12 & -0.44 &  5*  & a3116 \\
422320 & J084114.60+023503.6 & 0.0314 &   8 &  13 & 0.031408 & 223$\pm$   2 & 210 &  0.83$\pm$  0.09 &  0.70 &  15.8 &  9.54 & -0.02 &  1*  & a2934 \\
417392 & J084134.33+021119.8 & 0.0287 &  18 &  13 & 0.028710 & 192$\pm$   4 & 181 &  0.30$\pm$  0.04 &  0.38 &  11.5 &  9.03 &  0.15 &  1   & a2934 \\
345682 & J084242.49+015707.0 & 0.0255 &  10 &  12 & 0.025421 & 198$\pm$   4 & 187 &  0.88$\pm$  0.07 &  0.62 &  20.4 &  9.39 &  0.11 &  1*  & a2934 \\
417424 & J084333.16+021239.7 & 0.0299 &   8 &  13 & 0.029897 & 242$\pm$   5 & 228 &  0.80$\pm$  0.08 &  0.62 &  16.7 &  9.48 &  0.18 &  1   & a2934 \\
345754 & J084428.45+020350.2 & 0.0256 &  35 &  12 & 0.025518 & 261$\pm$  17 & 248 &  0.28$\pm$  0.05 &  0.37 &   9.5 &  8.89 & -1.36 &  1   & a3116 \\
517167 & J084438.72+022439.5 & 0.0299 &  19 &  13 & 0.029864 & 159$\pm$   7 & 148 &  0.20$\pm$  0.04 &  0.40 &   8.2 &  8.89 & -0.35 &  1*  & a3116 \\
422486 & J084443.28+023421.8 & 0.0261 &  20 &  10 & 0.026115 &  52$\pm$   5 &  46 &  0.18$\pm$  0.03 &  0.48 &  11.5 &  8.71 & -0.07 &  1*  & a3116 \\
517249 & J084612.24+022437.7 & 0.0281 &   8 &  13 & 0.028073 & 119$\pm$   7 & 110 &  0.51$\pm$  0.06 &  0.64 &  14.7 &  9.24 & -0.16 &  1*  & a3116 \\
517302 & J084654.29+023336.2 & 0.0287 &  10 &  13 & 0.028983 & 368$\pm$  20 & 351 &  1.81$\pm$  0.10 &  0.59 &  31.9 &  9.81 & -0.37 &  5*  & a2934 \\
323507 & J084808.41+013357.7 & 0.0399 &  25 &  13 & 0.039824 & 187$\pm$   1 & 173 &  0.26$\pm$  0.05 &  0.41 &   9.1 &  9.24 & -0.20 &  1   & a3116 \\
\hline	
\end{tabular}
\begin{flushleft}
Note. -- The full version of this table is available online.
\end{flushleft}
\end{table*}

\twocolumn

\section*{Appendix B: Notes on Individual Objects}\label{s_notes}

We list here notes for galaxies marked with an asterisk in the second-last column of Table~\ref{t_det}.
The galaxies are ordered by increasing SAMI number. In what follows, 
AA2 is the abbreviation for ALFALFA detection code 2.\\

\noindent
{\bf 14555 }  --    AA2; galaxy \about 3.5 arcmin SE has z=0.114.\\
{\bf 22839 }  --    blend emission from galaxy group, including large blue galaxy \about 2.5 arcmin SE (SDSS J115705.93+010732.1, $z=0.039517$), blue gal. \about 2 arcmin S 
                    (SDSS J115657.00+010657.0, $z=0.0393$) and small galaxy \about 3 arcmin S (SDSS J115654.22+010557.3, $z=0.039458$); edge-on disk \about 2.5 arcmin NE has $z=0.049$.\\
{\bf 22887 }  --    RFI spike at 1375 MHz (\about 9900 \kms); small galaxy \about 2 arcmin E has $z=0.079$.\\
{\bf 22932 }  --    AA2; disk galaxy \about 2.5 arcmin S has $z=0.064$.\\
{\bf 23337 }  --    galaxy \about 1 arcmin N has $z=0.078$; blue galaxy \about 3 arcmin SE has $z=0.066$.\\
{\bf 23669 }  --    RFI spike at 1392 MHz (\about 6150 \kms) visible in final spectrum; companion \about 2 arcmin S, SDSS J121035.31+010143.4 ($z=0.020115$, 6030 \kms) not detected. AA2.\\
{\bf 28860 }  --    blend with giant face-on spiral \about 3 arcmin SW, SDSS J142720.36+010133.1 (UGC 9254, $z=0.02569$, 7702 \kms).\\
{\bf 30377 }  --    high-frequency edge uncertain, systematic error.\\
{\bf 31452 }  --    small blue smudge \about 50 arcsec W without optical redshift, unlikely to significantly contaminate the signal; two blue galaxies within 4 arcmin in background ($z>0.06$).\\
{\bf 31509 }  --    RFI spike near 1382.5 MHz (\about 8300 \kms).\\
{\bf 31620 }  --    two background galaxies (small blue galaxy \about 30 arcsec S, SDSS J120155.54-010445.4, $z=0.117$ and small galaxy \about 3 arcmin NW, $z>0.11$); narrow RFI spike at 1392 MHz 
                    (within profile) not visible in spectrum.\\
{\bf 40765 }  --    low-frequency edge uncertain, systematic error.\\
{\bf 41164 }  --    RFI spike near 1392.5 MHz (\about 6100 \kms).\\
{\bf 56064 }  --    blue galaxy \about 2.5 arcmin NE has $z=0.071$.\\
{\bf 69740 }  --    AA2; galaxy \about 2.5 arcmin E has $z=0.058$.\\
{\bf 69844 }  --    blue fuzzy galaxy \about 10 arcsec W, no optical redshift.\\
{\bf 69864 }  --    galaxy \about 30 arcsec E has $z=0.095$.\\
{\bf 71382 }  --    RFI spike at 1388.8 MHz (\about 6800 \kms); small blue galaxy \about 1 arcmin SW has $z=0.073$.\\
{\bf 77373 }  --    two galaxies \about 2 arcmin N have $z=0.15$.\\
{\bf 79771 }  --    small face-on galaxy \about 30 arcsec NW, no optical redshift.\\
{\bf 84048 }  --    RFI spike near 1388.8 MHz (\about 6800 \kms).\\
{\bf 91926 }  --    polarization mismatch; two background galaxies (blue spiral \about 2 arcmin S, SDSS J141701.67+002837.4, $z=0.053$ and early-type \about 2.5 arcmin NE, SDSS J141712.85+003159.7, $z=0.077$).\\
{\bf 92770 }  --    galaxy \about 3 arcmin SW has $z=0.136$; galaxy \about 2.5 arcmin N has $z=0.118$.\\
{\bf 93167 }  --    AA2.\\
{\bf 99393 }  --    detected blue companion \about 2 arcmin N, SDSS J121037.24+010353.7 ($z=0.021841$, $cz=6548$ \kms).\\
{\bf 99511 }  --    blue galaxy \about 2.5 arcmin N has $z=0.083$.\\
{\bf 100192}  --    high-frequency edge uncertain, systematic error.\\
{\bf 105600}  --    confused/blend with edge-on disk \about 3 arcmin E, SDSS J141057.69+010208.6 ($z=0.025407$, 7617 \kms); large vel offset suggests that
                    this might be just one side of the \hi\ profile of the large companion.\\
{\bf 106049}  --    polarization mismatch.\\
{\bf 209279}  --    AA2.\\
{\bf 209708}  --    polarization mismatch.\\
{\bf 209743}  --    blue cloud \about 40 arcsec NE without optical redshift, possibly contaminating signal.\\
{\bf 220275}  --    polarization mismatch; AA2. Detected galaxy in OFF, SDSS J120842.76+012722.8 ($z=0.018547$, 5560 \kms). \\ 
{\bf 220371}  --    AA2.\\
{\bf 220578}  --    early-type galaxy \about 1 arcmin W has $z=0.129$ (NED); two other galaxies \about 1.5 arcmin NW, one has $z=0.130$, the other has no optical redshift, but likely in background. \\
{\bf 220750}  --    blue galaxy \about 20 arcsec NE, no optical redshift, possibly confused.\\
{\bf 227673}  --    AA2.\\
{\bf 228432}  --    blue galaxy \about 1 arcmin S, SDSS J142933.33+010616.0, possibly contaminating signal (no redshift in SDSS, NED has z=0.029490 from FoF??)\\
{\bf 230560}  --    galaxy \about 2.5 arcmin SW has $z=0.078$.\\
{\bf 230714}  --    two large galaxies \about 4 arcmin SE and \about 4 SW in foreground ($z=0.016$ and 0.019, respectively).\\
{\bf 239490}  --    AA2.\\
{\bf 251297}  --    AA2.\\
{\bf 251367}  --    blend with two blue galaxies: SDSS J143255.33+015209.0, \about 2 arcmin S ($z=0.030023$, 9001 \kms) and SDSS J143245.52+015238.8 \about 2.5 arcmin W ($z=0.029892$, 8961 \kms).\\
{\bf 273951}  --    blend with large galaxy 1 arcmin NW, SDSS J122339.84+011925.5 ($z=0.026436$, 7925 \kms); RFI spike at 1388.8 MHz (\about 6800 km/s). \\
{\bf 273952}  --    blue galaxy \about 4 arcmin SW, SDSS J122339.84+011925.5 ($z=0.026436$, 7925 \kms), partly detected?\\
{\bf 278684}  --    polarization mismatch.\\
{\bf 278741}  --    AA2; large spiral \about 3 arcmin S, SDSS J085339.27+004529.4 ($z=0.04116$, 12339 \kms), not detected.\\
{\bf 289116}  --    4 foreground galaxies within 4 arcmin ($z>0.022$).\\
{\bf 289429}  --    galaxy \about 3 arcmin W and \about 3 arcmin E in background ($z>0.11$).\\
{\bf 296685}  --    notice large edge-on disk \about 3 arcmin NW, SDSS J141110.08+012828.9, at the same redshift; however the \hi\ signal is so narrow that contamination is very unlikely.\\
{\bf 296847}  --    blend with blue companion \about 1 arcmin S, SDSS J141411.10+013006.4 (GAMA 296848 below). AA2.\\
{\bf 296848}  --    blend with blue companion \about 1 arcmin N, SDSS J141409.31+013111.2, $z=0.025927$ (GAMA 296847 above).\\
{\bf 301201}  --    AA2.\\
{\bf 302846}  --    blue disk \about 4 arcmin NE, SDSS J091843.59+012341.0, has $z=0.037$.\\
{\bf 320068}  --    blue gal. \about 2 arcmin NE has $z=0.025$; early type \about 1.5 arcmin SE has $z=0.100$.\\
{\bf 320281}  --    blend with large blue galaxy \about 2.5 arcsec NE, SDSS J143042.33+015253.2 ($z=0.033899$, 10163 \kms); small blue galaxy \about 1 arcmin N has $z=0.099$.\\
{\bf 345682}  --    notice blue blob \about 1.5 arcmin SE, SDSS J084245.27+015550.6 (no optical redshift).\\
{\bf 375531}  --    blend with blue galaxy \about 30 arcsec SE, SDSS J083852.35+012356.8 ($cz=8776$ \kms\ from NED, CGCG 004-098).\\
{\bf 375402}  --    large galaxy \about 1.5 arcmin NW, SDSS J083603.33+011947.1, no optical redshift.\\
{\bf 422289}  --    confused/blend with large disk galaxy \about 2 arcmin E, SDSS J084114.59+023503.6 ($z=0.031423$, 9420 \kms).\\
{\bf 422320}  --    AA2.\\
{\bf 422486}  --    RFI spike at 1388.8 MHz (\about 6800 \kms).\\
{\bf 517167}  --    galaxy \about 1 arcmin E has $z=0.077$.\\
{\bf 517249}  --    AA2.\\
{\bf 517302}  --    blend with blue companion \about 1 arcmin NW, SDSS J084651.22+023410.9 ($z=0.029603$, 8875 \kms); also notice large early-type galaxy \about 2.5 arcmin W, 
                    SDSS J084643.96+023214.7 ($z=0.028128$, 8433 \kms).\\
{\bf 517594}  --    AA2. Possibly contaminated/blend with blue galaxy \about 40 arcsec NE, SDSS J085201.74+023256.6 (no optical redshift).\\
{\bf 522166}  --    blend with small blue companion \about 20 arcsec W, SDSS J085210.96+025709.8 ($z=0.028986$, 8690 \kms).\\
{\bf 534759}  --    galaxy merger 20 arcsec SW has $z=0.079$; few other galaxies within 3 arcmin in background.\\
{\bf 537163}  --    blue galaxy \about 40 arcsec E has $z=0.081$; small blue galaxy \about 1 arcmin SE has no optical redshift; 
                    2 red galaxies \about 2 arcmin NW and \about 2.5 arcmin W also in background ($z>0.1$).\\
{\bf 537187}  --    superimposed on red galaxy, no optical redshift; blue companion \about 3 arcmin E, SDSS J121803.44-005447.0 ($z=0.021048$), some contamination possible.\\
{\bf 537417}  --    blue galaxy 2 arcmin S has $z=0.040$.\\
{\bf 543812}  --    disturbed morphology; small galaxy 3 arcmin E in foreground (SDSS J141224.85-004955.2, $z=0.026017$).\\
{\bf 551202}  --    small reddish galaxy \about 2.5 arcmin NE, SDSS J091734.93-003457.5 ($z=0.016785$, 5032 \kms) very unlikely to significantly contaminate signal.\\
{\bf 558887}  --    blend with larger, blue companion \about 2.5 arcmin S, SDSS J113733.94-003056.0 ($z=0.028865$, 8654 \kms).\\
{\bf 567624}  --    small blue galaxy \about 3 arcmin E in background (SDSS J141026.38-003518.7, $z=0.053725$).\\
{\bf 592466}  --    background blue galaxy \about 2 arcmin S (SDSS J141221.01-000927.6, $z=0.081$).\\
{\bf 599839}  --    low-frequency edge uncertain, systematic error. Large early-type companion \about 1 arcmin SW not detected (SDSS J085109.79+001512.9, $z=0.040914$, 12266 \kms); 
                    disk galaxy \about 2.5 arcmin N has $z=0.052$.\\
{\bf 600014}  --    AA2. Large companion \about 3 arcmin NE, SDSS J085358.68+002100.1 ($z=0.028157$, 8441 \kms); galaxy \about 10 arcsec S has $z=0.167$;
                    two other gals \about 2 arcmin NW and 3 arcmin NW in background.\\
{\bf 600030}  --    small blue companion \about 3 arcmin SW, SDSS J085353.63+001815.2 ($z=0.028331$, 8493 \kms), might contaminate the signal; small red galaxy S of companion 
                    and one \about 2.5 arcmin W in background ($z=0.167$ and 0.078 respectively).\\
{\bf 610474}  --    high-frequency edge uncertain, systematic error; two gals within \about 2 arcmin in background ($z=0.104$ and 0.105).\\
{\bf 610997}  --    AA2.\\
{\bf 618071}  --    AA2. \\
{\bf 618935}  --    3 background galaxies within 4 arcmin ($z>0.09$).\\
{\bf 619046}  --    early-type galaxy \about 2.5 arcmin S has $z=0.054$.\\
{\bf 619094}  --    merger; blend with companion \about 1.5 arcmin NW, SDSS J143208.40+002440.1 ($z=0.034636$, 10384 \kms); galaxy \about 2.5 arcmin NW has $z=0.053$. AA2.\\
{\bf 619098}  --    AA2. Blue galaxy \about 3 arcmin SE in background (SDSS J143221.79+001041.2, $z=0.054790$ from NED).\\
{\bf 622858}  --    polarization mismatch.\\
{\bf 623144}  --    AA2.\\

\section*{Appendix C: Galaxies with wrong inclination estimates or with highly disturbed \Ha\ kinematics}\label{s_incl}

\begin{figure*}
\begin{center}
\includegraphics[width=17.5cm]{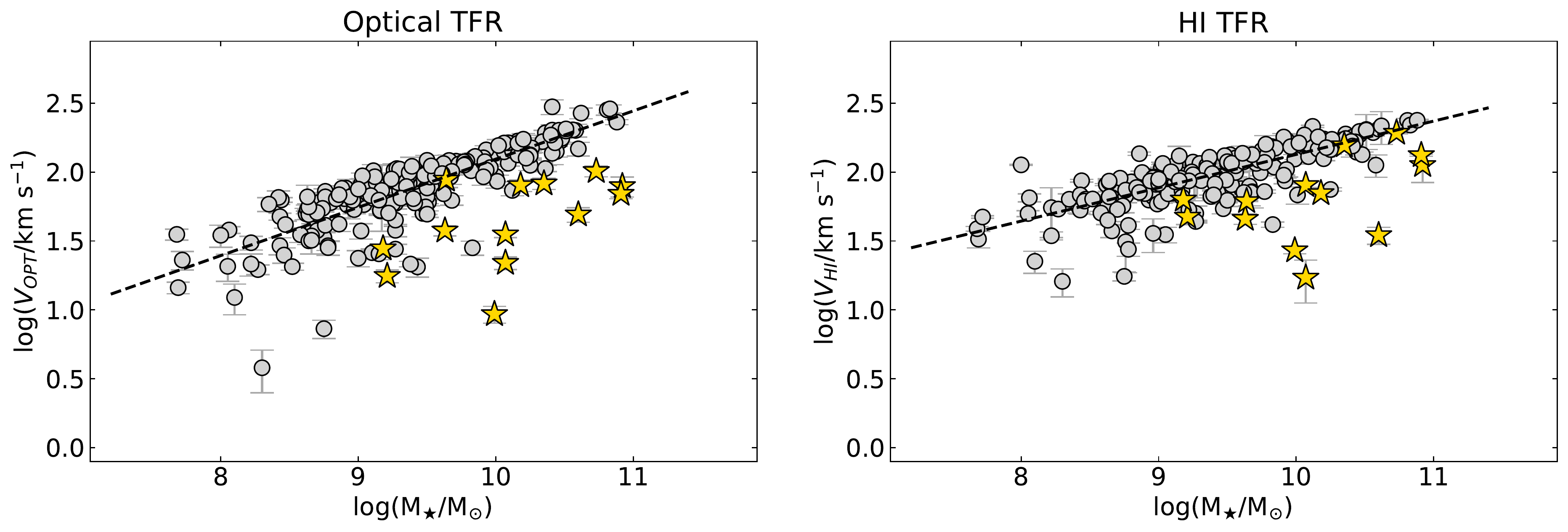}
\caption{Optical (left) and \hi\ (right) stellar mass TFRs and linear fits for the kinematic, pruned SAMI-\hi\ sample, reproduced from the bottom panels of Fig.~\ref{fig_tfr} (gray circles). Yellow stars indicate additional galaxies with incorrect inclination estimates that were discarded after visual inspection (see  Fig.~\ref{fig_excl_SDSS}), consistently with our $i<40$\deg\ selection cut.
}
\label{fig_excl}
\end{center}
\end{figure*}

\begin{figure*}
\begin{center}
\includegraphics[width=14cm]{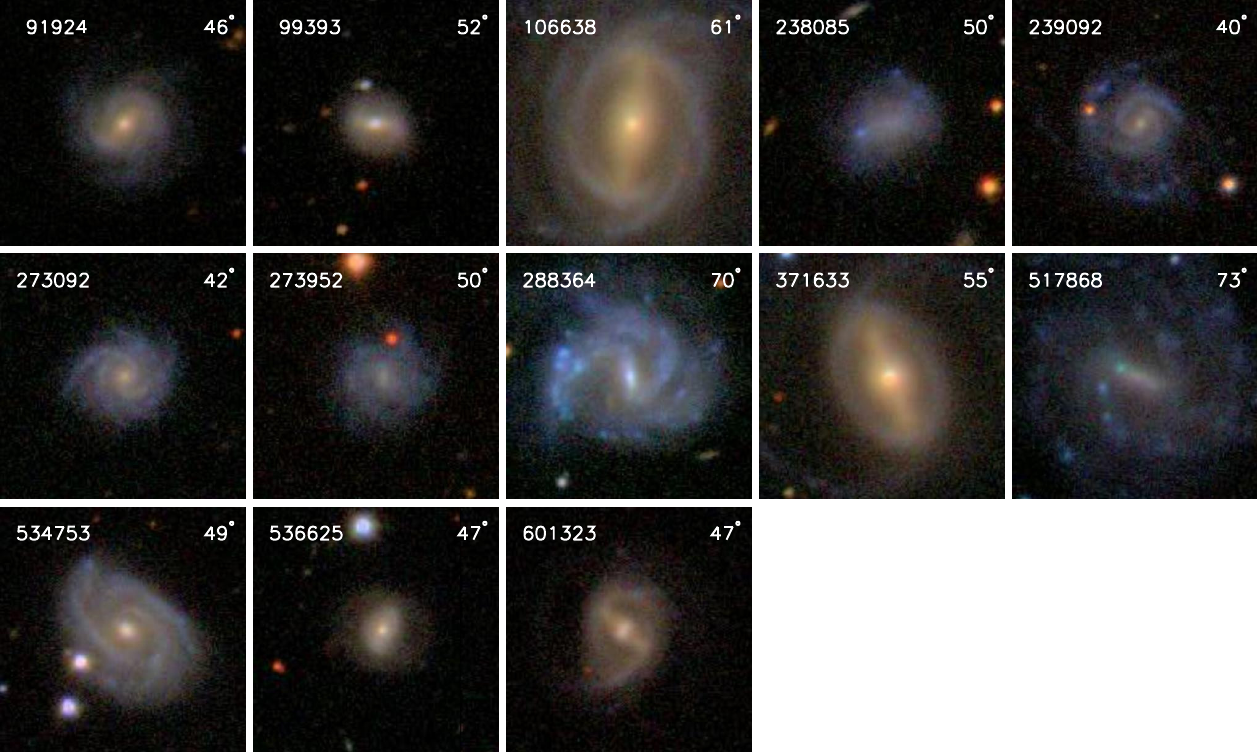}
\caption{SDSS postage stamp images (1 arcmin square) of the galaxies marked as yellow stars in Fig.~\ref{fig_excl}, with SAMI identifier and inclination noted on the top left and right corners, respectively.}
\label{fig_excl_SDSS}
\end{center}
\end{figure*}

We give here some more details on galaxies with incorrect inclinations or with kinematically disturbed \Ha\ velocity fields that were identified by visual inspection. Fig.~\ref{fig_excl} shows the positions on the stellar TFRs of 13 galaxies with incorrect inclinations, which are for the most part outliers in these plots. As can be seen from their SDSS images in Fig.~\ref{fig_excl_SDSS}, several of these systems have strong bars, and their actual inclinations are smaller than 40\deg, hence are not included in the SAMI-\hi\ kinematic, pruned sample (see \S~\ref{s_tfr}).

\begin{figure*}
\begin{center}
\includegraphics[width=16cm]{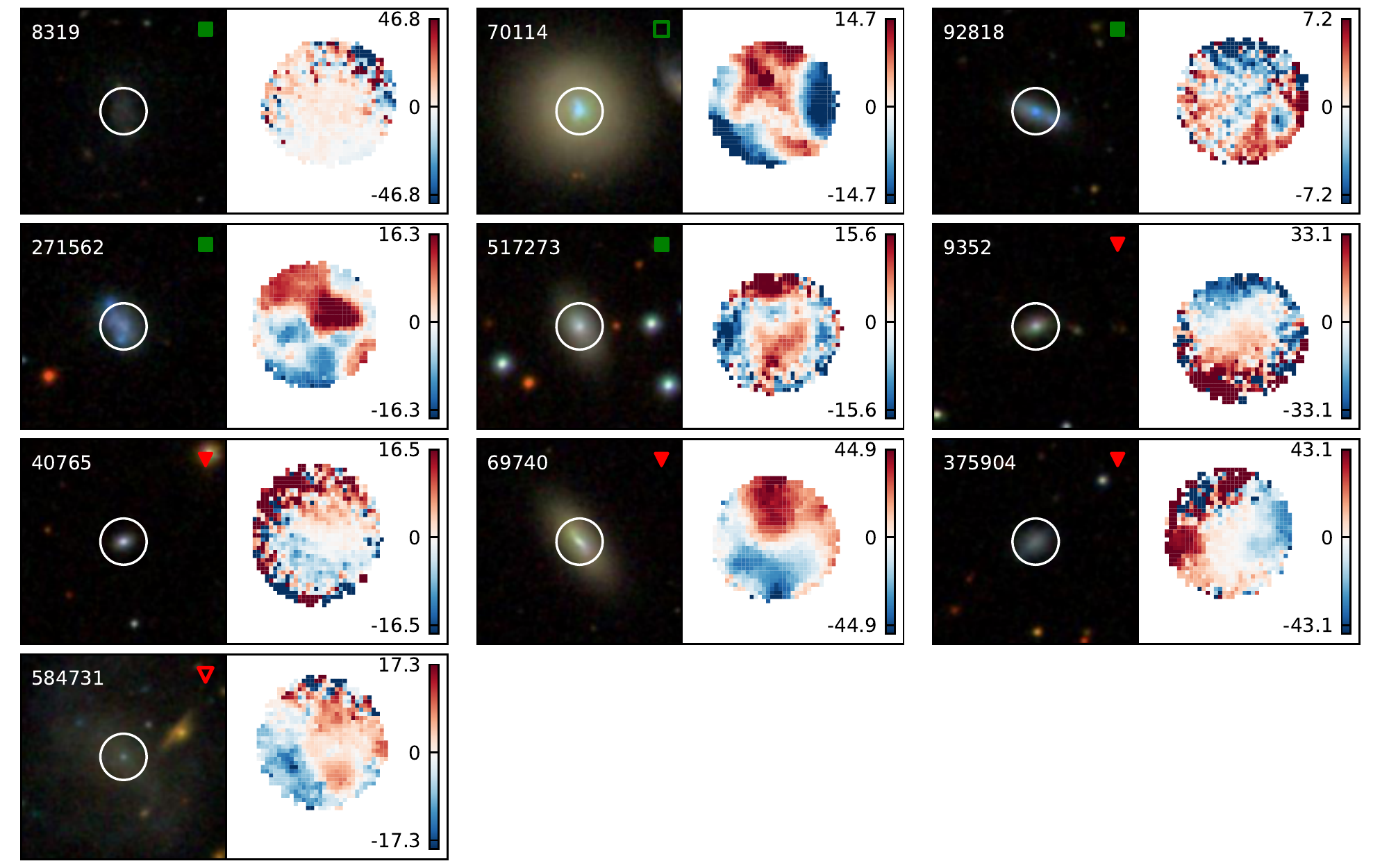}
\caption{SDSS postage stamp images (1 arcmin square) and SAMI \Ha\ velocity fields of kinematically disturbed systems marked with green and red symbols in Fig.~\ref{fig_vratio}a. On the SDSS images, SAMI identifiers and symbols used in Fig.~\ref{fig_vratio}a are noted on the top-left and top-right corners, respectively; white circles show the \about 15 arcsec field-of-view of SAMI. Color bars next to the maps are labelled by \Ha\ line-of-sight velocity in \kms.}
\label{fig_outliers}
\end{center}
\end{figure*}

Fig.~\ref{fig_outliers} shows SDSS images and SAMI \Ha\ velocity fields of 10 galaxies with complex/disturbed kinematics, as discussed in \S~\ref{s_vel} and \S~\ref{s_tfr}.

\bsp	
\label{lastpage}
\end{document}